\documentclass[aps,pre,showpacs,twocolumn]{revtex4}
\usepackage{amsfonts}
\usepackage{amsmath}
\usepackage{amsthm}
\usepackage{graphics}
\usepackage{graphicx}
\usepackage{subfigure}
\usepackage{amssymb}
\usepackage{color}
\usepackage[T1]{fontenc}
\begin{document}

\title{Axial dipolar dynamo action in the Taylor-Green vortex}

\author{Giorgio Krstulovic$^{1,2}$, Gentien Thorner$^1$, Julien-Piera Vest$^{1}$, Stephan Fauve$^{1}$ and Marc Brachet$^{1}$}
\affiliation{$^{1}$Laboratoire de Physique Statistique de l'Ecole Normale 
Sup{\'e}rieure, \\
associ{\'e} au CNRS et aux Universit{\'e}s Paris VI et VII,
24 Rue Lhomond, 75231 Paris, France.}
\affiliation{$^{2}$Laboratoire Cassiop\'ee, Observatoire de la C\^ote d'Azur, CNRS, Universit\'e de Nice Sophia-Antipolis, Bd. de l'Observatoire, 06300 Nice, France
}
\date{\today}
\pacs{47.65.-d, 47.20.Ky, 52.65.Kj}

\begin{abstract}
We present a numerical study of the magnetic field generated by the Taylor-Green vortex. We show that periodic boundary conditions can be used to mimic realistic boundary conditions by prescribing the symmetries of the velocity and magnetic fields. This gives insight in some problems of central interest for dynamos: the possible effect of velocity fluctuations on the dynamo threshold, the role of boundary conditions on the threshold and on the geometry of the magnetic field generated by dynamo action. In particular, we show that an axial dipolar dynamo similar to the one observed in a recent experiment can be obtained with an appropriate choice of the symmetries of the magnetic field.
The nonlinear saturation is studied and a simple model explaining the magnetic Prandtl number dependence of the super/sub critical nature of the dynamo transition is given.
\end{abstract}

\maketitle
\bigskip


\section{Introduction}
The generation of magnetic field by the flow of an electrically conducting fluid has been mostly studied to understand the magnetic fields of planets and stars \cite{moffatt1978}. 
The increase in computing power has allowed the study of these fluid dynamos in almost realistic three-dimensional geometries although far from the parameter range relevant for astrophysical or geophysical flows or even laboratory experiments \cite{roberts2000}.

Three successful laboratory experiments have been performed so far: fluid dynamos have been observed with an helical Ponomarenko-type flow \cite{gailitis2001} or with a array of helical flows with the same helicity \cite{stieglitz2001}. In these two first experiments, the flow lines were strongly constrained by the boundaries, thus restricting turbulent fluctuations to small scales, and the observed dynamos were found in very good agreement with the ones obtained by the action of the mean flow alone. A third example of fluid dynamo has been provided in the laboratory by the VKS experiment, i.e., a von K{\'a}rm{\'a}n (VK) flow of liquid sodium \cite{monchaux2007}. In contrast to the previous dynamo experiments, the geometry of the generated magnetic field strongly differs from the predictions made by kinematic dynamo codes using the mean flow alone. The mean magnetic field has to leading order a dipolar structure with its axis along the axis of rotation of the propellers that drive the flow \cite{monchaux2009}, whereas kinematic dynamo codes using the mean flow alone predict an equatorial dipole. Thus, the VKS dynamo is not generated by the mean flow alone. It has been proposed that non axisymmetric fluctuations related to the propellers driving the flow play a crucial role in the generation of an axial dipolar magnetic field  \cite{petrelis2007} and this mechanism has been illustrated by a kinematic dynamo simulation using a non axisymmetric model flow in a cylindrical geometry \cite{gissinger2009}. 

Two different choices can be made for the numerical simulations of dynamos. One possibility is to consider a flow confined in a finite domain and to use boundary conditions as realistic as possible. Another one is to use periodic boundary conditions on the velocity and magnetic fields. 
In the later case, the Taylor-Green (TG) flow has been widely studied. The TG vortex is a standard turbulent flow often used in numerical computations \cite{BRACHET:1983p4817} that is closely related to the experimentally studied VK swirling flow \cite{DouadyPRL,Fauve1993,Maurer1994}.
The relation between the VK flow and the TG vortex is a similarity in overall geometry: a shear layer between two counter-rotating eddies. The TG vortex, however, is periodic with impermeable free-slip boundaries (present as mirror symmetries) while the experimental flow takes place between two counter-rotating coaxial impellers and is confined inside a cylindrical container. The TG vortex also obeys a number of additional rotational symmetries.

Dynamo action in the TG flow was studied some time ago \cite{Nore:1997p5767} and it was found that the most unstable growing magnetic mode is equatorial (perpendicular to the axis of rotation) and breaks the additional rotational symmetries. Thus enforcing all the geometric symmetries of the TG flow was found to be non-favorable to dynamo action. Since that time all numerical studies of dynamo action in the TG vortex were performed using general periodic codes (not enforcing any of the symmetries). The generated magnetic field was always found to be equatorial \cite{ponty2005,ponty2007,yadav2010}.
                     
The new idea that we explore in the present paper is to enforce the mirror symmetries of the TG flow (that confine the flow) but not enforce the rotational symmetries of the magnetic field (that forbid the appearance of magnetic dipoles). With mirror symmetries enforced, the corresponding boundary conditions on the magnetic field can be related to either electrically insulating \footnote{More precisely corresponding to magnetic field normal to the boundary, see the discussion at the end of Sec. \ref{sec:DecompSym}} or perfectly conducting boundaries \cite{Lee_2008}. In the present case (without the rotational symmetries) there are $2^3=8$ possibilities for the magnetic boundaries conditions. 

The paper is organized as follows: after recalling the MHD equations and the TG forcing, we present the symmetries of the TG flow in section \ref{Sec:TGSymetries}. We show that there exists only one choice of mirror symmetry for the velocity field that is compatible with the equations of motion. It corresponds to free-slip boundary conditions on each boundary of the cube $[0, \pi]^3$. In contrast, in each directions two independent choices exist for the magnetic field, either corresponding to an electrically insulating boundary, or to a perfectly conducting boundary. We show that we can always restrict the kinematic dynamo study in a symmetric subspace and that there are six different cases to consider. In section \ref{sec:Symdyn} we first shortly describe the numerical procedures and explain how the growth rates of the different dynamo modes are determined. Then, we consider how the imposed symmetries on the velocity field affect the dynamo threshold. We show that in the absence of constraints, the flow can spontaneously break the TG forcing symmetries, thus generating turbulent fluctuations that inhibit dynamo action although involving larger kinetic energy. In section \ref{Sec:BCeffect}, we consider the different symmetry constraints for the magnetic field and their respective dynamo efficiency. We show that the lower threshold is obtained when lateral boundaries are of different nature (two conducting and two insulating). In addition, the lowest threshold is obtained when the upper and lower ones are insulating. In this case, the threshold value and the neutral mode geometry correspond to the ones already reported in a general periodic code without symmetry constraints \cite{Nore:1997p5767}. We observe that the geometry of the dynamo modes strongly depends on the symmetry constraints that are chosen. With all boundaries electrically insulating, we obtain the generation of an axial dipolar magnetic field, as the one observed in the VKS experiment. 
The nonlinear saturation of this axial dipole is studied in section \ref{sec:saturation} where we also present a simple model explaining the super/sub critical nature of the dynamo transition when the magnetic Prandtl number is varied.
Discussion and conclusions are given in section \ref{sec:conclusion}.

\section{Theoretical preliminaries \label{Sec:TGSymetries}}
\subsection{MHD equations and Taylor-Green forcing}\label{sec:MHD}

The magnetohydrodynamics (MHD) equations for an unit density fluid and an incompressible flow read in terms of the velocity $\bf v$ and magnetic induction $\bf b$ (in units of Alfv\`en velocity):
\begin{eqnarray}
\partial_t {\bf v} + {\bf v} \cdot {\bf \nabla v} &=& -\nabla P + {\bf j} \times {\bf b}  + \nu \Delta {\bf  v}+\bf{f}\label{EQ:MHDv}\\
\partial_t {\bf b} &=& \nabla \times ({\bf v} \times {\bf b})  + \eta \Delta {\bf b}\label{EQ:MHDb}
\end{eqnarray}
together with $\nabla \cdot \bf v = 0 = \nabla \cdot \bf b$; $\bf j=\nabla \times \bf b$ is the current density, $P$ is the pressure, $\bf{f}$ is the mechanical forcing, $\nu$ is the kinematic viscosity and $\eta$ is the magnetic diffusivity.
The total energy is
\begin{eqnarray}
E(t)&=&E_v(t)+E_b(t)\label{Fig:Energies}\\
E_v(t)=\frac{1}{2} \langle {{\bf v}}^2\rangle &&E_b(t)=\frac{1}{2} \langle{\bf b}^2 \rangle \label{eq:EnergiesDef}
\end{eqnarray}
(where $E_v(t)$ is the kinetic energy, $E_b(t)$ is the magnetic energy and $\langle \ \rangle$ stands for spatial average over the domain).
The total energy is conserved by these equations in the ideal case ($\nu=\eta=0$ and ${\bf f}={\bf 0}$). 

Considering a flow that is $2 \pi$-periodic in all spatial dimensions, the kinematic Reynolds number ${\rm Re}$ and the magnetic Reynolds number  ${\rm Re_m}$ are defined as
\begin{equation}
{\rm Re}=\frac{L v_{\rm rms}}{\nu},\hspace{.5cm} {\rm Re_m}=\frac{L v_{\rm rms}}{\eta}\label{Eq:defReynolds}
\end{equation}
where the root-mean square velocity is $v_{\rm rms}=\sqrt{2 E_v/3}$ and the characteristic length $L$ is defined by 
\begin{equation}
L=2\pi\frac{\sum_{k} k^{-1}E_v(k,t)}{\sum_{k} E_v(k,t)\,dk},
\end{equation}
where  the kinetic energy spectrum $E_v(k,t)$ (such that $E_v(t)=\sum_{k}E_v(k,t)$) is obtained by summing
${\frac1 2}|{\bf \hat u}({\bf k'},t)|^2 \,$  on the spherical shells $k-1/2\leq |{\bf k'}|<k+1/2$
(${\bf \hat u}({\bf k})$ is
the Fourier transform of the velocity).
Analogously, the magnetic energy spectrum is denoted by $E_b(k,t)$ and verifies $E_b(t)=\sum_{k} E_b(k,t)$.
The magnetic Prandtl number of the fluid is given by ${\rm P_m} = \nu / \eta =  {\rm Re_m}/{\rm Re}$.

We now turn to the definition of the external driving volumic force ${\bf f}$ in \eqref{EQ:MHDv} that balances the energy dissipation and allows to reach a statistically stationary state that is needed to sustain dynamo action.
Following reference \cite{Nore:1997p5767}, we force the system by setting in \eqref{EQ:MHDv} 
\begin{equation}\label{Eq:defForce}
{\bf f}=f(t) {\bf v^{\rm TG}},
\end{equation}
where ${\bf v^{\rm TG}}$ is the Taylor-Green \cite{TaylorGreen} initial data
\begin{equation}
{\bf v^{\rm TG}}=(\sin(x) \cos(y) \cos(z),-\cos(x) \sin(y)\cos(z), 0), \label{eq:VTG}
\end{equation}
and $f(t)$ is determined by imposing that
the projection of $\bf v$ on the mode ${\bf v^{\rm TG}}$ is fixed at all times to its initial value.

In the Navier-Stokes problem (Eq. \eqref{EQ:MHDv}, with $ {\bf j} ={\bf b}=0$) is well-known \cite{BRACHET:1983p4817} that a number of the symmetries of ${\bf v^{\rm TG}}$ are compatible with the equations of motion. They are, first, 
rotational symmetries of angle $\pi$ around the axis $(x=z=\pi/2)$ and $(y=z=\pi/2)$; and of angle $\pi/2$ around the axis $(x=y=\pi/2)$. 
Another set of symmetries corresponds to planes of mirror symmetry: $x=0,\pi$, $y=0,\pi$ and $z=0,\pi$
that form the sides of the so-called impermeable box which confines the flow.
However, it is important to note that the TG symmetries of the solution can be \emph {spontaneously broken}, in the sense that a small non-symmetric component of the initial data can grow and eventually break the symmetry of the solution \cite{Nore:1997p5767}. The symmetry breaking of the TG flow will be numerically studied in Sec.\ref{Sec:SymBreaking}.

\subsection{Mirror symmetries revisited}

The main new idea of the present paper is to enforce the mirror symmetries of the TG flow (that confine the flow) but not enforce the rotational symmetries of the magnetic field (that forbid the appearance of magnetic dipoles). With mirror symmetries enforced, the corresponding boundary conditions on the magnetic field can be related to either electrically insulating or perfectly conducting boundaries \cite{Lee_2008} (see the discussion at the end of Sec. \ref{sec:DecompSym}). 
In the present case (without the rotational symmetries) it will turn out that there are new (mixed) possibilities for the magnetic boundaries conditions. 
We now proceed to explore all the TG mirror symmetries that are dynamically compatible with the full MHD equations (\ref{EQ:MHDv}-\ref{EQ:MHDb}).

The MHD equations (\ref{EQ:MHDv}-\ref{EQ:MHDb}) are invariant under reflections with respect to a plane. 
We now turn to considerations pertaining to this propriety.
Let us first define the standard reflection transformation
of the vector 
${\bf r}=(r_1,r_2,r_3) \equiv(x,y,z)$ 
about the plane 
$r_\alpha=0$ 
by ${\bf S}^\alpha$
with
\begin{eqnarray}
{\bf S}^1 (r_1,r_2,r_3)= (-r_1,r_2,r_3) , \\
{\bf S}^2 (r_1,r_2,r_3)= (r_1,-r_2,r_3) , \\
{\bf S}^3 (r_1,r_2,r_3)= (r_1,r_2,-r_3) .
\end{eqnarray}
Note that ${\bf S}^\alpha$ is its own inverse.

The action of the reflection operation on a vector field ${\bf h}({\bf r})$ is defined by
\begin{equation}\label{Eq:defRefleProy}
{\bf R}^\alpha({\bf h}({\bf r}))= {\bf S}^\alpha {\bf h}({\bf S}^\alpha {\bf r})
\end{equation}
which explicitly reads, in the case of the the $z=0$ plane:
\begin{equation}
{\bf R}^z \begin{pmatrix}   
      h_x(x,y,z) \\
      h_y(x,y,z) \\
      h_z(x,y,z) 
  \end{pmatrix}
=\begin{pmatrix}   
      h_x(x,y,-z) \\
      h_y(x,y,-z) \\
     -h_z(x,y,-z) 
  \end{pmatrix}.
\end{equation}
We will say that the vector field ${\bf h}$ has a mirror-symmetry with respect to the plane $r_\alpha$ if it is even  (${\bf R}^\alpha{\bf h}=-{\bf h}$) or odd (${\bf R}^\alpha{\bf h}={\bf h}$) under transformation \eqref{Eq:defRefleProy}. Note that if a vector field is even (odd) with respect to a plane, then it is perpendicular (parallel) to that plane. When a vector field that is odd (or even) with respect to simultaneous transformations with respect to all planes, it is said that it transforms as a  \emph{vector} (or \emph{pseudo-vector}).
Note that the Taylor-Green vortex is odd with respect to all planes: ${\bf R}^\alpha{\bf v}^{\rm TG}={\bf v}^{\rm TG}$ for all $\alpha$. 

The following properties are true for any fields ${\bf g}$ and ${\bf h}$
\begin{eqnarray}
{\bf R}^\alpha ({\bf\nabla} \times {\bf h})&=&-{\bf \nabla}\times({\bf R}^\alpha{\bf h})\label{Eq:PropApp1} ,\\
{\bf R}^\alpha({\bf g}\times{\bf h})&=&-({\bf R}^\alpha{\bf g})\times ({\bf R}^\alpha{\bf h})\label{Eq:PropApp2} ,\\
R^\alpha \triangle{\bf g}&=&{\triangle}({\bf R}^\alpha{\bf g})\label{Eq:PropApp3}.
\end{eqnarray}

From Eq.\eqref{Eq:PropApp1} it is directly observed that if ${\bf h}$ is even with respect to a plane then ${\bf\nabla} \times {\bf h}$ is odd with respect to that plane and vice versa. 


We now turn to study the mirror symmetries of the MHD equations. Applying ${\bf R}^\alpha$ to (\ref{EQ:MHDv}-\ref{EQ:MHDb}) and writing ${\bf v} \cdot \nabla {\bf v}=(\nabla \times {\bf v}) \times {\bf v}+  \nabla (\frac{1}{2}\bf {\bf v}^2)$, it directly follows from Eqs.(\ref{Eq:PropApp1},\ref{Eq:PropApp2},\ref{Eq:PropApp3}) that ${\bf \bar{v}}={\bf R}^\alpha{\bf v}$ and ${\bf \bar{b}}={\bf R}^\alpha{\bf b}$ satisfy
\begin{eqnarray}
\nonumber\partial_t {\bf \bar{v}} + (\nabla \times {\bf \bar{v}}) \times \bf {\bf \bar{v}}&=& -\nabla( \bar{P} +\frac{1}{2}\bf {\bf \bar{v}}^2)+ (\nabla \bf {\bf \bar{b}}) \times \bf {\bf \bar{b}}  \\
&&+ \nu \Delta \bf {\bf \bar{v}}+{\bf R}^\alpha\bf{f}\label{EQ:MHDvR}\\
\partial_t {\bf \bar{b}} &=& \nabla \times ({\bf \bar{v}} \times {\bf \bar{b}})  + \eta \Delta {\bf \bar{b}}\label{EQ:MHDbR},
\end{eqnarray}
with $ \bar{P}({\bf r})=P({\bf S}^\alpha {\bf r})$. Therefore the only symmetry of the velocity that is compatible with the MHD equations (\ref{EQ:MHDv}-\ref{EQ:MHDb}) is ${\bf R}^\alpha\bf{f}={\bf f}$ and ${\bf \bar{v}}={\bf v}$. 
This can be easily understood with the following simple geometrical argument based on vortex rings that are known to have a self-induced motion along their axis. Suppose that there exist a symmetry plane containing a vortex ring, for such a configuration the vorticity is parallel to the plane. The related velocity field is thus an even field with respect to that plane. It is immediately noticed that the symmetry will not be preserved by the Navier-Stokes equation: the vortex ring will leave the plane under it's self-induced motion thus breaking the symmetry. Another possibility is to place the vortex ring perpendicular to the plane, making the velocity parallel, therefore odd with respect to that plane. The self-induced motion will then respect the symmetry.

Note however that the magnetic field must be symmetric but it can be either even or odd . Therefore, for a mirror symmetric velocity field, the magnetic field has two possible symmetries for each plane that are compatible with the MHD equations. We will see below that the two possibilities correspond to insulating (I) and  perfectly conducting (C) magnetic boundary conditions (of the free-slip type) for the even and odd cases respectively. Note that, as the the fields are defined in a $2\pi$-periodical box, the planes $x=\pi,y=\pi$ and $z=\pi$ are also mirror symmetry planes with the same symmetries than $x=0,y=0$ and $z=0$. In the following, depending on the symmetries of the magnetic field, we will refer to these planes as walls of type I (even) or C (odd). With this definition, the symmetry planes of the Taylor-Green vortex are of type CCC (vorticity is perpendicular to the box $[0,\pi]^3$).

\subsection{Projection on symmetric magnetic fields \label{Sec:ProjTheo}}

Let us now define the projectors into symmetric functions with respect to the planes $r_i=0$ and $r_i=\pi$,
\begin{equation}
{\bf Q}_{s}^\alpha=\frac{1}{2}(I-s {\bf R}^\alpha),
\end{equation}
 where $\alpha$ stands for $x,y,z$, ${\bf R}^\alpha$ is defined by Eq.\eqref{Eq:defRefleProy} and $s=\pm1$. Note that by construction ${\bf R}^\alpha({\bf Q}_{s}^\alpha{\bf h})=-s {\bf Q}_{s}^\alpha{\bf h}$ and therefore $Q_{x_i}^s{\bf h}$ is an even vector (if $s=+1$) or odd vector ($s=-1$) with respect to the planes $r_\alpha=0$ and $r_\alpha=\pi$. 
 
 As the above discussion about the two possible magnetic mirror symmetries was done for each plane $r_\alpha=0$ independently, we now define the projector over the most general mirror-symmetric magnetic fields corresponding to a different choice for each plane. 
 \begin{equation}
 {\bf P}^{\vec{ s}}={\bf Q}^{1}_{s_1}{\bf Q}^{2}_{s_2}{\bf Q}^{3}_{s_3}\label{Eq:DefFullProj},
 \end{equation}
 with $\vec{ s}=(s_x,s_y,s_z)\in\{-1,1\}^3$. Note that by construction 
  \begin{equation}
 {\bf P}^{(-1,-1,-1)}{\bf v^{\rm TG}}={\bf v^{\rm TG}} \label{Eq:VTGProj}.
  \end{equation}

Let us now consider a completely mirror-symmetric $2\pi$-periodical velocity field. With this choice ${\bf v}$ is confined into the impermeable box $[0,\pi]^3$: there is no flow crossing the boundaries. This mimics the rigid boundary conditions of experiments that are also impermeable. We now explain how the different kinds of magnetic symmetries are related to perfectly conducting and insulating boundary conditions. If the magnetic field is even with respect to one of the mirror-symmetry planes, we call this plane a \emph{insulating} (I) wall because the current ${\bf j}=\nabla \times{\bf b}$ is parallel to (or contained in) the wall \cite{Lee_2008}. Analogously if ${\bf b}$ is odd the wall is called \emph{conducting} (the current is perpendicular to (or crosses) the wall.

We now show that any $2\pi$-periodical magnetic field can be decomposed in a finite sum of mixed symmetrical-periodical vector and pseudo-vector fields that correspond respectively to insulating and perfectly conducting magnetic boundary conditions. Thus the complete study of the influence of symmetries (boundary conditions) on the TG dynamo threshold is reduced to a finite number of possible cases.

Let $\Omega$ be the space of $2\pi$-periodic functions, and  $\Omega_{\vec{ s}}= {\bf P}^{\vec{ s}}\Omega$ the projection of $\Omega$ over the subspaces with mirror symmetries given by $\vec{ s}$. For example if ${\bf b}\in \Omega_{1,1,-1}$, then ${\bf j}=\nabla\times{\bf b}$ is normal to the planes $z=0$ and $z=\pi$ and parallel to the other ones.  Therefore $\Omega_{1,1,-1}$  is the $2\pi$-periodical function space such that the planes $x=0,x=\pi,y=0,y=\pi$ are \emph{insulating} and the planes $z=0,z=\pi$ are \emph{conducting}.

It can be demonstrated by straightforward (but tedious) computations that
 \begin{eqnarray}
 {\bf P}^{\vec{ s}_1} {\bf P}^{\vec{s}_2}&=&I\delta_{\vec{ s_1},\vec{ s_2}}\\
  \sum\limits_{\vec{ s} } {\bf P}^{\vec{ s}}&=&I.\label{Eq:IdentDecompSym}
 \end{eqnarray}
We thus have that $\Omega$ decomposes as the direct sum
 \begin{equation}
 \Omega=\bigoplus_{\vec{ s}}\Omega_{\vec{ s}},\label{Eq:DirectSum}
 \end{equation}
 and therefore the general-periodic function space is then decomposed in eight mirror-symmetric periodical-function spaces.
 For the sake of clarity, in the following we will denote each space $\Omega_{\vec{ s}},$ by its respective type of walls. For instance the case $\Omega_{1,1,-1}$ is labeled  IIC.
  
\subsection{Decomposition of kinematic dynamo on symmetry classes}\label{sec:DecompSym}

The kinematic dynamo is the linear instability of the MHD equations linearized about a given velocity field ${\bf v}_{\rm s}$. 
For a stationary ${\bf v}_{\rm s}$, the growth rate $\sigma_b$ of the magnetic field is given by the eigenvalue problem $\mathcal{L}{\bf b}=\sigma_b {\bf b}$ with 
\begin{equation}
\mathcal{L}{\bf b}=\nabla \times ({\bf v}_{\rm s} \times {\bf b})  + \eta \Delta {\bf b}\label{EQ:Dynamo}.
\end{equation}
Observe that the eigenvalue $\sigma_b$ depends on $\eta$ and thus is a function of ${\rm Re_{m}}$ (see Eq.\eqref{Eq:defReynolds}). The dynamo threshold is defined by the critical magnetic Reynolds number ${\rm Re_{m}^{crit}}$ such that $\sigma_b({\rm Re_{m}^{crit}})=0$. Note that this number is not necessary unique due to windows of instability \cite{ponty2007}. In this work we refer to ${\rm Re_{m}^{crit}}$ as the smallest critical magnetic Reynolds number.
 
As a consequence of the symmetry invariance of the MHD Eqs.(\ref{EQ:MHDv}-\ref{EQ:MHDb}), it is straightforfard to show that the symmetry projectors $P_{\vec{ s}}$ commute with $\mathcal{L}$. Therefore if ${\bf b_0}$ is an eigenvector associated to the eigenvalue $\sigma_b^0$, by the decomposition of the identity \eqref{Eq:IdentDecompSym}, there exist one symmetry $\vec{s_0}$ such that $ P_{\vec{ s_0}}{\bf b_0}$ is also a eigenvector associated to $\sigma_b^0$. We thus have for the threshold of general-periodic dynamo
\begin{equation}
{\rm Re_{m}^{crit}}=\min_{\vec{s}}{{\rm {  \{Re_{m}^{crit}}^{\vec{s}}}\}},\label{Eq:TheoRemc}
\end{equation}
 where ${\rm Re_{m}^{crit}}^{\vec{s}}$ is the critical magnetic Reynolds number of the linear problem \eqref{EQ:Dynamo} restricted to $\Omega_{\vec{s}}$. In other words, a bifurcating mode of the magnetic field has a certain mirror-symmetry.  It is thus natural to study the dynamo bifurcation for each symmetry by restricting \eqref{EQ:Dynamo} to each $\Omega_{\vec{s}}$. This can be easily done if each mirror-symmetry is imposed to the magnetic field by applying the projectors $P_{\vec{s}}$.
 
Since the forcing \eqref{Eq:defForce} is invariant by rotation of $\pi/2$ along the $x=y= \pi/2$ axis,
the spaces ICI and ICC are respectively equivalent to CII and CIC. There are thus only six independent cases to study. The symmetries III, CCC, IIC, CCI, ICI and ICC will be individually studied in Sec.\ref{Sec:BCeffect}.

Note that we are dealing throughout this paper with periodic fields, with physical boundaries being replaced by symmetry planes.
In contrast, in the case of a physical fluid confined into a finite domain a magnetic field at an insulating boundary should be matched with a potential field outside the flow domain. The case that we call ``insulating'' also mimics a boundary with a magnetic permeability much larger than that of the fluid, {\it i.e.} liquid sodium inside an iron vessel. In that case matching with a potential field is not required and the magnetic field is perpendicular to the boundaries.

\section{Symmetries of the velocity field and dynamo threshold} \label{sec:Symdyn}

 \subsection{Numerical procedures and determination of growth rates \label{Sec:NumMethods}}

Numerical solutions of Eqs.(\ref{EQ:MHDv}-\ref{EQ:MHDb}) are efficiently produced using the pseudospectral general-periodic code  GHOST \cite{Mininni:GHOST}, that is dealiased by spherical spectral  truncation using the $2/3$-rule \cite{Got-Ors}. Thus a run with resolution $N^3$ has a maximum wavenumber $k_{\rm max}=N/3$. Resolutions used in this works vary from $64^3$ to $256^3$. The equations are evolved in time using a second-order Runge-Kutta method, and the code is fully parallelized with the message passing interface MPI library.  We implemented into GHOST both the constant velocity forcing \eqref{Eq:defForce} and the projectors \eqref{Eq:DefFullProj}.

The TG vortex \eqref{eq:VTG} is used as initial data for Eq.\eqref{EQ:MHDv}, eventually adding a small non-symmetric random part when studying symmetry breaking (see below section \ref{Sec:SymBreaking}). 
A small, spectrally band-limited random seed of given magnetic energy is used as initial data for Eq.\eqref{EQ:MHDb}.
 
\begin{figure}
\begin{center}
\includegraphics[width=8.5cm]{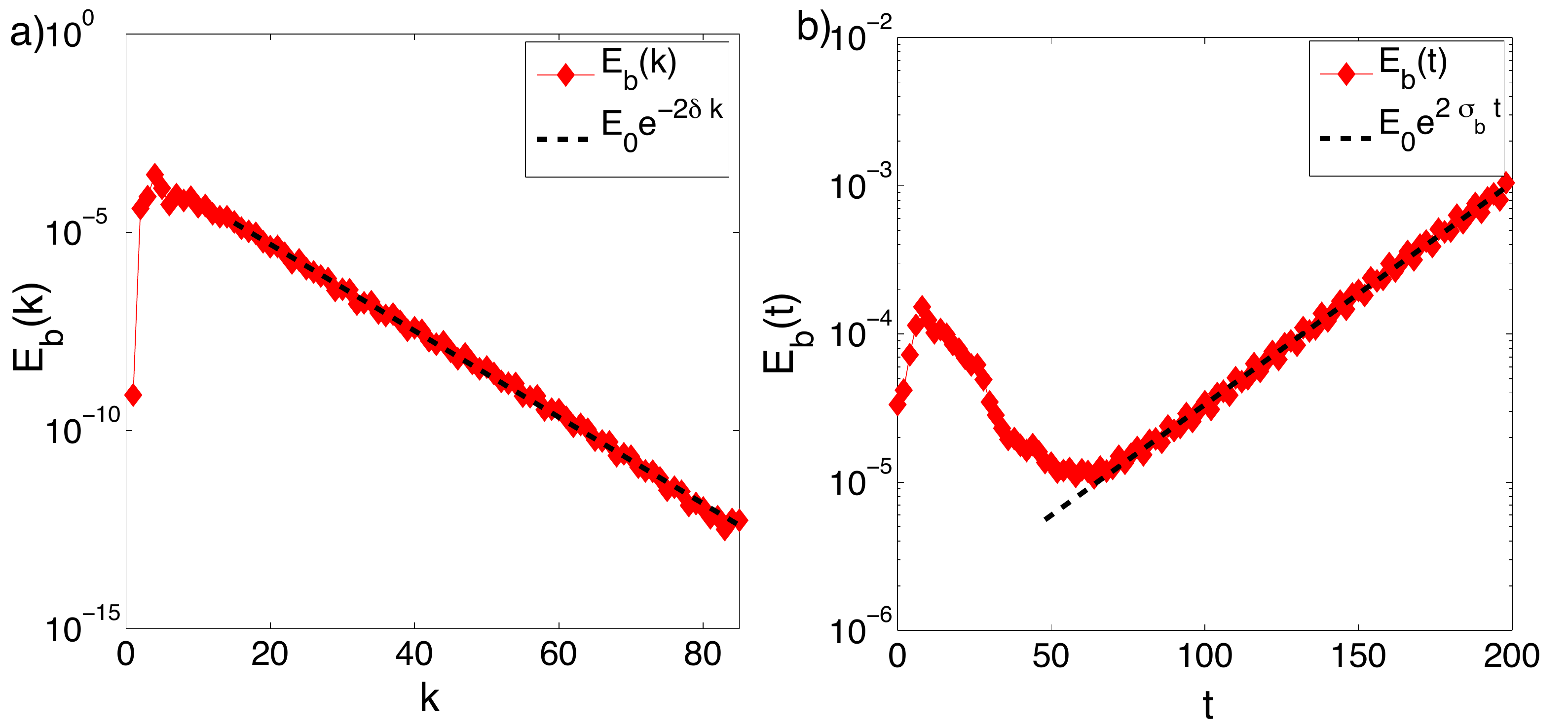}
\caption{a) Magnetic energy spectrum of a dynamo run: case III with ${\rm Re=30}$, ${\rm Re_m=80}$. b) Corresponding temporal evolution of magnetic energy \eqref{Fig:Energies}.  The fits used to determine $\sigma_b=0.017$ and $\delta k_{\rm max}=10.72$ (for the present simulation) are displayed as (dashed) straight lines of Figs. a) and b).}
\label{Fig:energies}
\end{center}
\end{figure}

To correctly resolve the MHD Eqs.(\ref{EQ:MHDv}-\ref{EQ:MHDb}) spectrally, a fast decay at large $k$ (faster than algebraic) of the energy spectrum is required. This condition (called spectral convergence) is quantitatively determined by fitting the exponential decay of the energy spectra $E_v(k,t)$ and $E_b(k,t)$ by a law of the form $C  e^{- 2 \delta k}$ that amounts to a simple Lin-Log linear regression. The value of  $\delta k_{\rm max}$ furnishes a measure of spectral convergence.
For instance,  Fig. \ref{Fig:energies} shows a numerical simulation where the magnetic field is well resolved,  as apparent on the Lin-Log plot of $E_b(k)$. A fit of the data in Fig.\ref{Fig:energies}.a leads to $\delta_b k_{\rm max}\ge10$. 
In all the runs presented in this work,  we always have $\delta k_{\rm max}>2.5$, with this condition we ensure that the fields are well resolved and that there is no spurious numerical effect on the observed dynamo instabilities.

Once the correct resolution of MHD equations is ensured the next step is to observe the behavior of the magnetic energy $E_b(t)$. Within the linear theory of kinetic dynamo $\mathcal{L}{\bf b}=\sigma_b {\bf b}$ (see Eq.\eqref{EQ:Dynamo}) we expect an exponential behavior of $E_b(t)$ with a growth rate $\sigma_b$. For each run  the magnetic energy is thus fitted with a law of the form $E_b(t)=C e^{\sigma_b t}$ to determine $\sigma_b$.  Figure \ref{Fig:energies}.b shows a typical temporal evolution of  $E_b(t)$ and the corresponding fit gives the value $\sigma_b=0.017$.
The data in Fig.\ref{Fig:energies} thus correspond to a typical dynamo action run.

 \subsection{Symmetry breakings of velocity field\label{Sec:SymBreaking}}

As already stated at the end of Sec.\ref{sec:MHD}, the TG symmetries of the solution to Eq.\eqref{EQ:MHDv} can be {\emph spontaneously broken}, in the sense that a small non-symmetric component of the initial data can grow to order one and completely break the symmetry of the solution. Two types of symmetry breaking will be considered in this section: first confinement breaking and second (with confinement enforced) the breaking of the rotational symmetry of angle $\pi/2$ around the axis $(x=y=\pi/2)$. We now turn to a numerical study of these points.

\subsubsection{Spontaneous confinement breaking}
\begin{figure}
\begin{center}
\includegraphics[width=8.5cm]{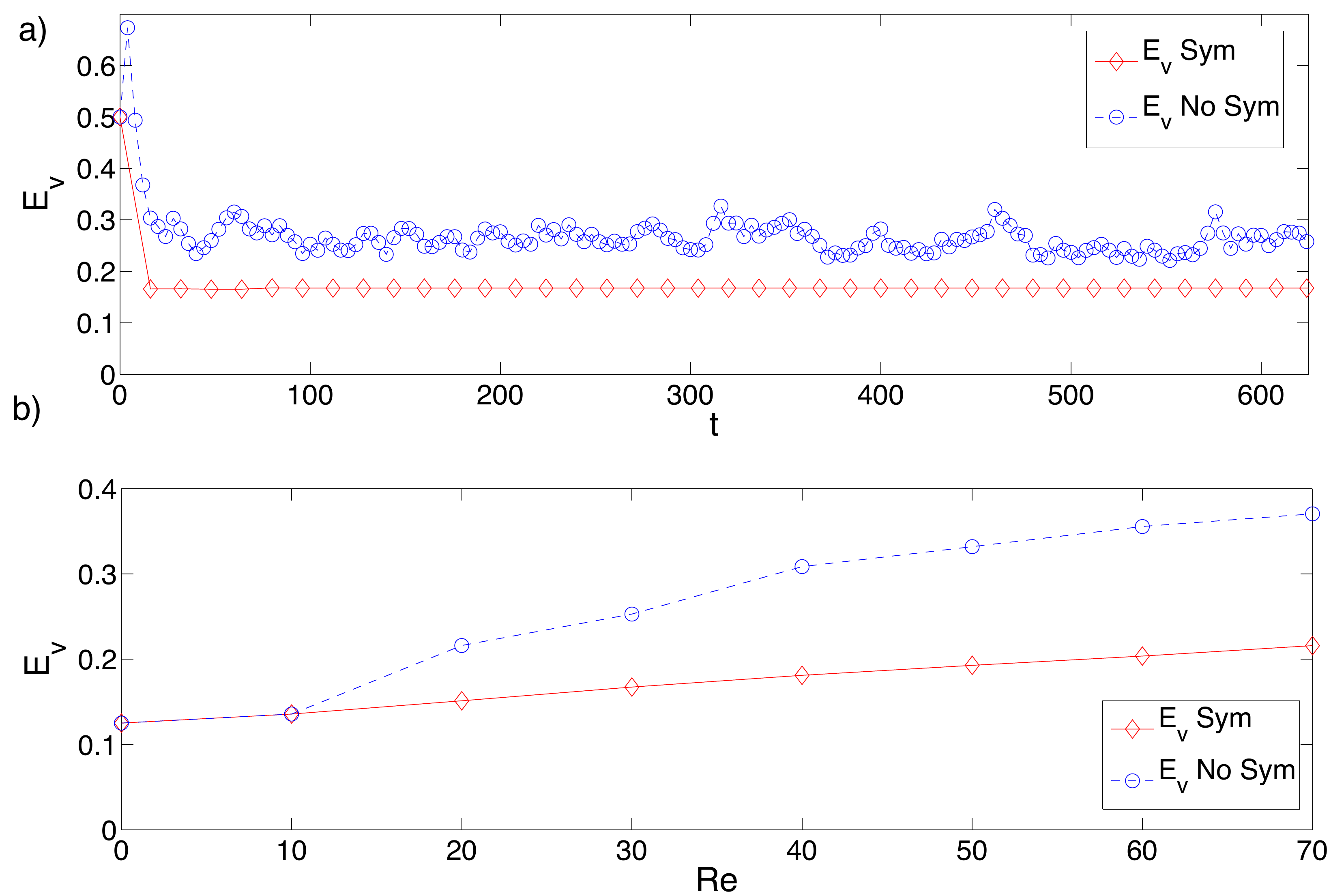}
\caption{(Color online) a) Temporal evolution of kinetic energy with (red) and without (blue) confinement \eqref{Eq:VTGProj} imposed at ${\rm Re}=30$. b) Dependence of kinetic energy  on the Reynolds number for the symmetric solid (red) line and non-symmetric dashed (blue) line runs (time-averaged over statistically stationary values).}
\label{Fig:symmetrybreaking}
\end{center}
\end{figure}

To study the spontaneous breaking of the mirror symmetries of the TG flow (that confine the flow) we compare runs performed with and without projecting the velocity field after each timestep using \eqref{Eq:VTGProj}.
The time dependence of the kinetic energy at ${\rm Re}=30$ is shown in \ref{Fig:symmetrybreaking}.a, where it is seen that (when non enforced) the mirror symmetry is broken with an increase of the kinetic energy $E_k(t)$ due to fluctuations of the velocity field. 
Mirror (confinement) symmetry breaking is seen to take place around  ${\rm Re}=10$ in Fig.\ref{Fig:symmetrybreaking}.b, where the dependence of the kinetic energy (time-averaged over statistically stationary values) on ${\rm Re}$ is displayed. 

\subsubsection{Spontaneous $\pi/2$ rotation breaking}\label{sec:symbreakpis2}

\begin{figure}
\begin{center}
\includegraphics[width=8cm]{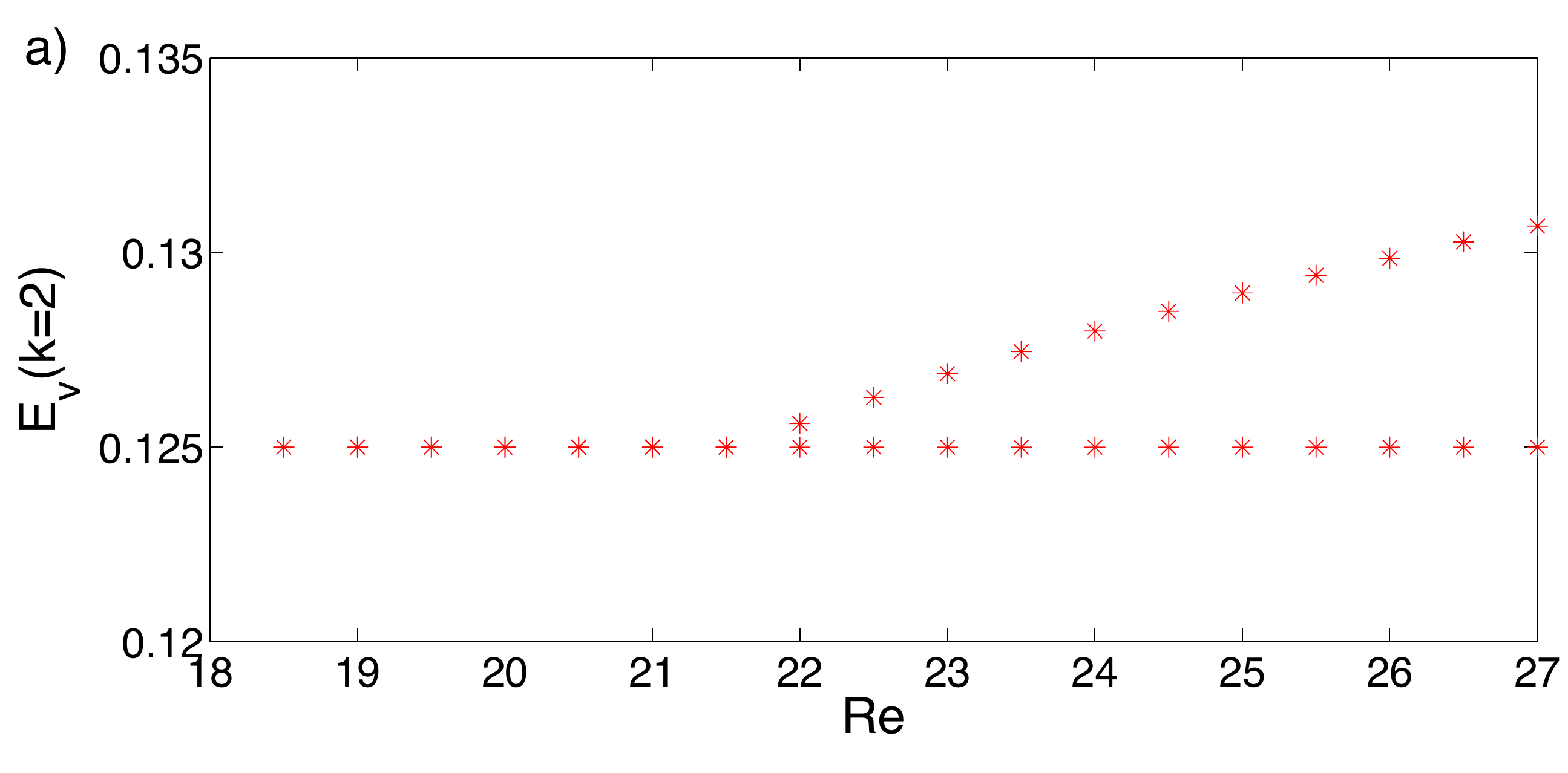}
\includegraphics[width=4.25cm]{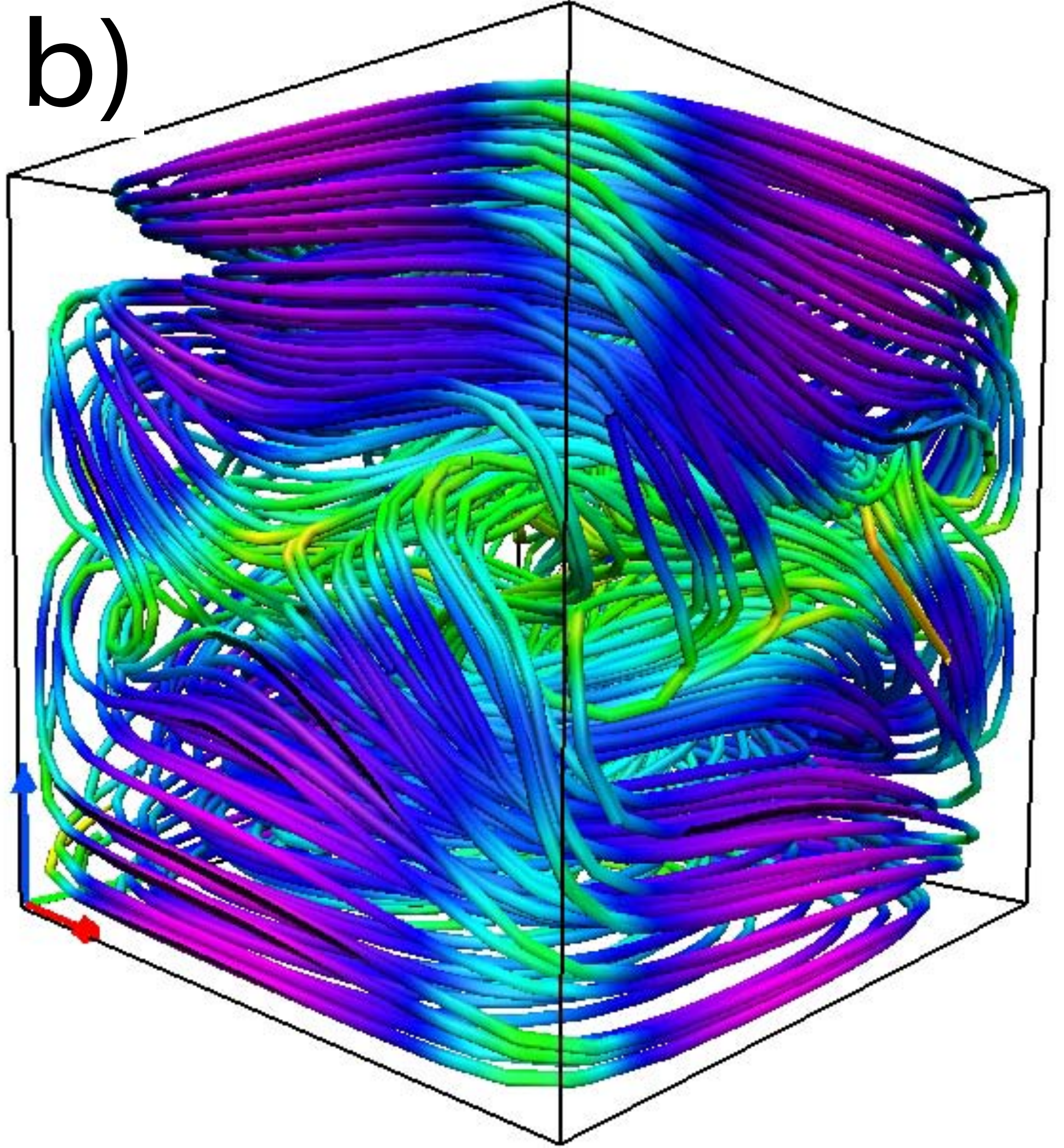}
\includegraphics[width=4.25cm]{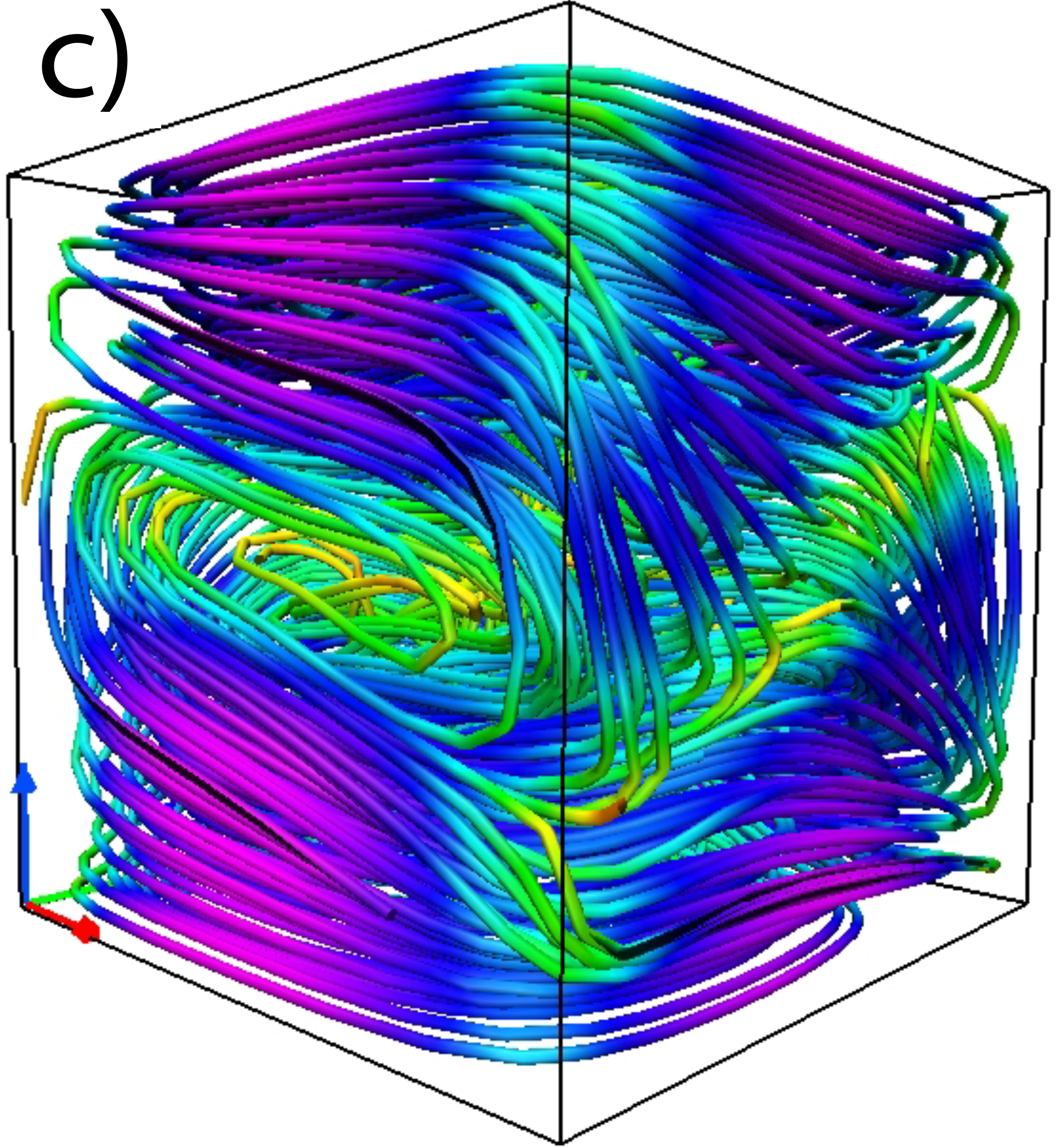}
\caption{(Color online) a) Bifurcation diagram: kinetic energy $E_v(k=2)$ as function of ${\rm Re}$. A pitchfork bifurcation is clearly present ($E_v(k=2)$ is quadratic in the bifurcating mode ${\bf v^{\rm PF}}$). b-c) Visualization of Taylor-green stationary states at  ${\rm Re}=30$: non-bifurcated (b) and bifurcated (c) velocity fields. Streamlines are colored accordingly to the magnitude of the velocity field, varying form magenta (low intensity zones) to yellow (high intensity zones).}
\label{Fig:TGvis}
\end{center}
\end{figure}

We now turn to the numerical study of another bifurcation of the steady-state velocity field that takes place when confinement \eqref{Eq:VTGProj} is enforced.  Stable and unstable steady states are followed by making use of Newton's method, the linear equations being solved by the stabilized  bi-congugate gradient algorithm \cite{gb}, see  \cite{newton} and references therein. The result of such computations is displayed in Fig.\ref{Fig:TGvis}.a, where it is apparent that a pitchfork bifurcation is present at critical  Reynolds number  ${\rm Re^c}=22$.
Physical $3D$ visualizations of the velocity fields are made using VAPOR \footnote{http://www.vapor.ucar.edu} and presented in Fig.\ref{Fig:TGvis} b-c.
By computing the difference between the stable and unstable branches obtained near the bifurcation by Newton's method the bifurcating mode can be numerically obtained. The dominant Fourier component ${\bf v^{\rm PF}}$ (PF standing for pitchfork) is found in this way to be 
\begin{equation}
{\bf v^{\rm PF}}=\begin{pmatrix}   
      \sin(x) \cos(y) \cos(z) \\
      \cos(x) \sin(y)\cos(z)\\
      -2\cos(x)\cos(y)\sin(z)  \label{EQ:vpf}
  \end{pmatrix}.
\end{equation}
It is straightforward to check that $\bf v^{\rm PF}$ is odd under $\pi/2$ rotation around the axis $(x=y=\pi/2)$. 
As the equations and the non-bifurcated states are both invariant by this transformation the bifurcation is a pitchfork (see Fig.\ref{Fig:TGvis}.a) that breaks the $\pi/2$ rotational invariance around the axis $(x=y=\pi/2)$.

\subsection{Effect of velocity fluctuations on magnetic energy growth rates}

\begin{figure}
\begin{center}
\includegraphics[width=8.5cm]{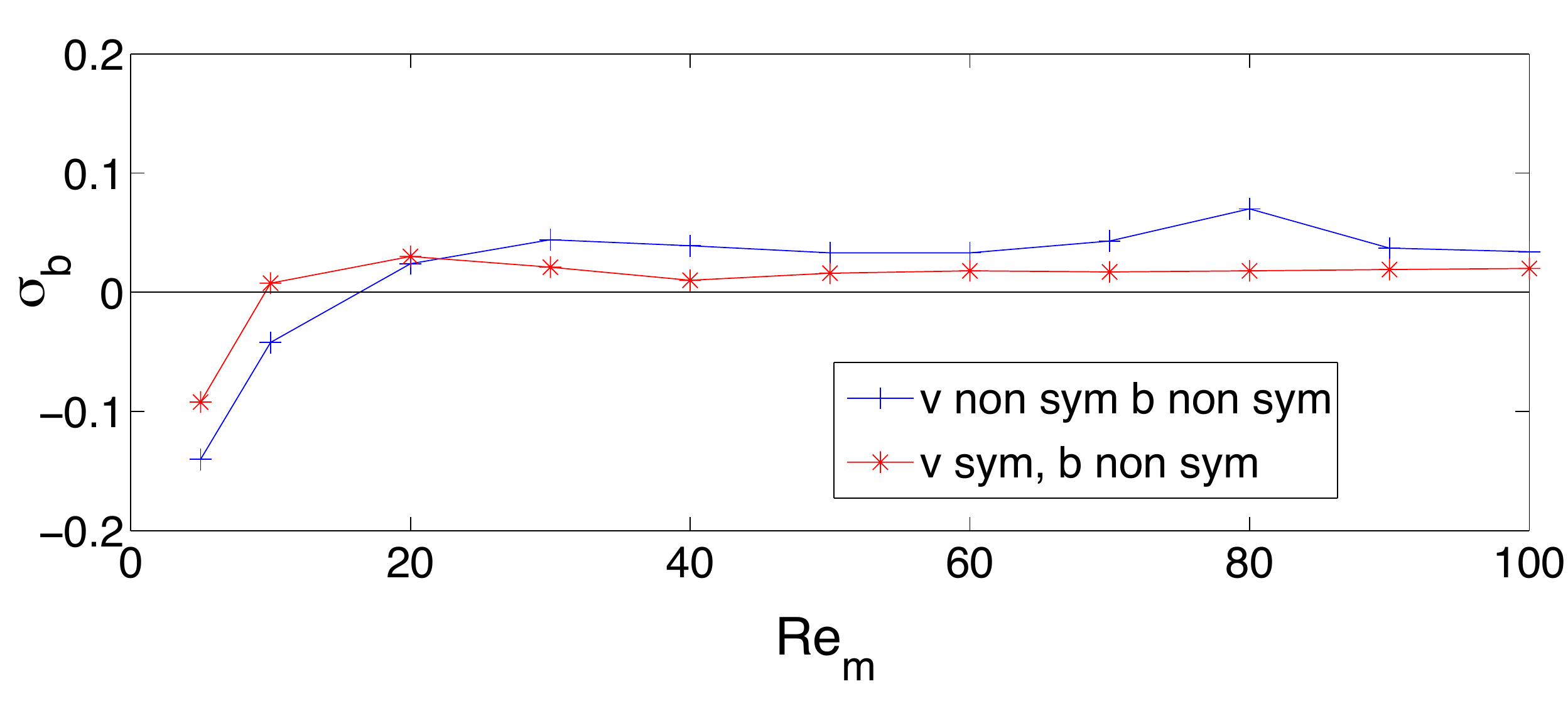}
\caption{(Color online) Dynamo growth rates $\sigma$ corresponding to the symmetric and non symmetric velocity fields of Fig.\ref{Fig:symmetrybreaking}a as function of the magnetic Reynolds number ${\rm Re_m}$. Blue crosses: ${\bf v}$ and ${\bf b}$ non symmetric, Red stars: ${\bf v}$ symmetric and ${\bf b}$ non symmetric.}
\label{Fig:sigmaVsymETnonsym}
\end{center}
\end{figure}

The symmetries of the confined velocity fields reduce the fluctuations of the velocity field, as observed in Fig.\ref{Fig:symmetrybreaking}.
Using the methods outlined in section \ref{Sec:NumMethods} to determine the magnetic growth rates, 
we observe that this reduction enhances the magnetic instability and reduces the critical magnetic Reynolds number. This is apparent in Fig.\ref{Fig:sigmaVsymETnonsym} where the growth rates  for the non-symmetric  and symmetric velocity field are compared as functions of ${\rm Re_m}$ at fixed kinematic Reynolds number ${\rm Re}=30$.

\section{Effect of magnetic boundary conditions\label{Sec:BCeffect}}

\subsection{Magnetic boundary conditions and dynamo threshold}

\begin{figure}
\begin{center}
\includegraphics[width=8.5cm]{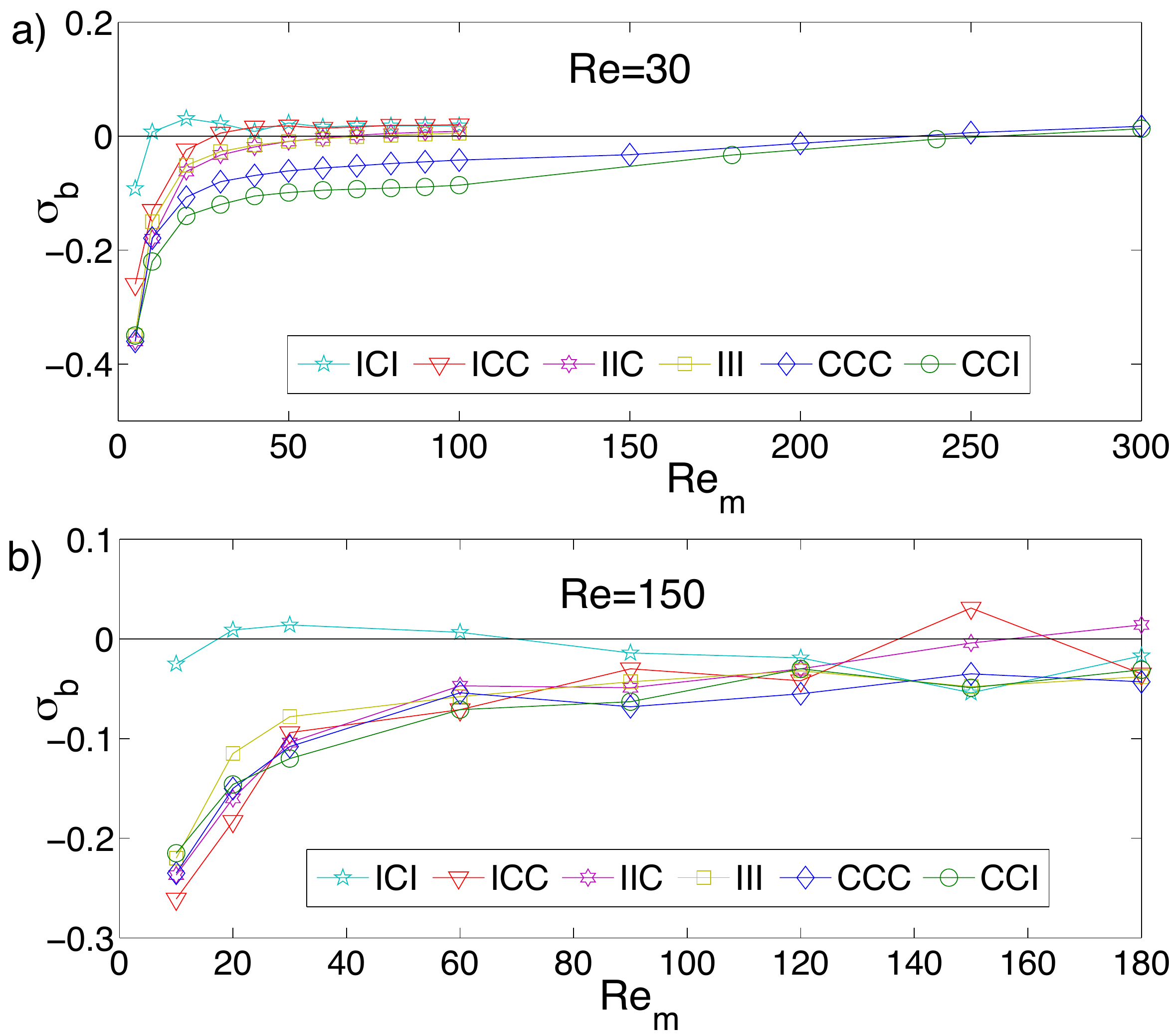}
\caption{(Color online) Dynamo growth rates $\sigma_b$ (symmetric velocity field) as function of the magnetic Reynolds number ${\rm Re_m}$  for the $6$ possible symmetries of the magnetic field at kinetic Reynolds number ${\rm Re}=30$ (a) and ${\rm Re}=150$ (b). }
\label{Fig:sigmaBsym30}
\end{center}
\end{figure}

We now focus on a confined velocity field (with mirror symmetries imposed) and study the influence of the different kinds of magnetic boundary conditions on the dynamo threshold. A number of runs have been performed at kinematic Reynolds numbers ${\rm Re}=30$, ${\rm Re}=150$ and magnetic Reynolds numbers varying between ${\rm Re_m}=10-300$ with resolutions of $64^3-256^3$ grid points. For each run, the growth rate is measured and the whole ensemble of data is presented in Fig. \ref{Fig:sigmaBsym30}. This is our main quantitative result on growth rates. 

By performing a linear interpolation of $\sigma_b$ we obtain the critical magnetic Reynolds number ${\rm Re_m^{crit}}$ that are given in Table \ref{Table:CriticRem} for ${\rm Re}=30$.
\begin{table}[h]
\begin{tabular}{| c || c | c | c | c | c | c |}
  \hline
  Case & ICI & ICC & IIC & III & CCC & CCI  \\ \hline
  ${\rm Re_m^c}$ & $9$ & $26$ & $66$ & $73$ & $231$ & $254$ \\ 
    \hline
 \end{tabular}
 \caption{Critical magnetic Reynolds number for the different walls and symmetric and non-symmetric cases at ${\rm Re}=30$. Values obtained by linear fit of $\sigma_b$. \label{Table:CriticRem}}
  \end{table}
Observe that critical magnetic Reynolds numbers vary from  ${\rm Re_m^{crit}}\sim10$ to  ${\rm Re_m^{crit}}\sim250$ for the different kind of walls. 

The most favorable cases correspond to mixed insulating-conducting lateral walls (ICI and ICC). These configurations allow for a magnetic field crossing the box in the direction perpendicular to the insulating walls and current crossing in the other direction. The less favorables cases turn out to correspond to the lateral perfectly conducting walls (CCC and CCI). Note that, see the discussion at the end of Sec.\ref{sec:DecompSym}, the condition we call insulating corresponds to a magnetic field perpendicular to the boundary. This case can be achieved experimentally by using a ferromagnetic boundary. 

Fig. \ref{Fig:sigmaBsym30} also displays the growth rate for ${\rm Re}=150$. At this Reynolds number, the velocity field is turbulent and the fluctuations increase the dynamo threshold. Observe that the case ICI is still the most unstable.

\begin{figure}
\begin{center}
\includegraphics[width=8.5cm]{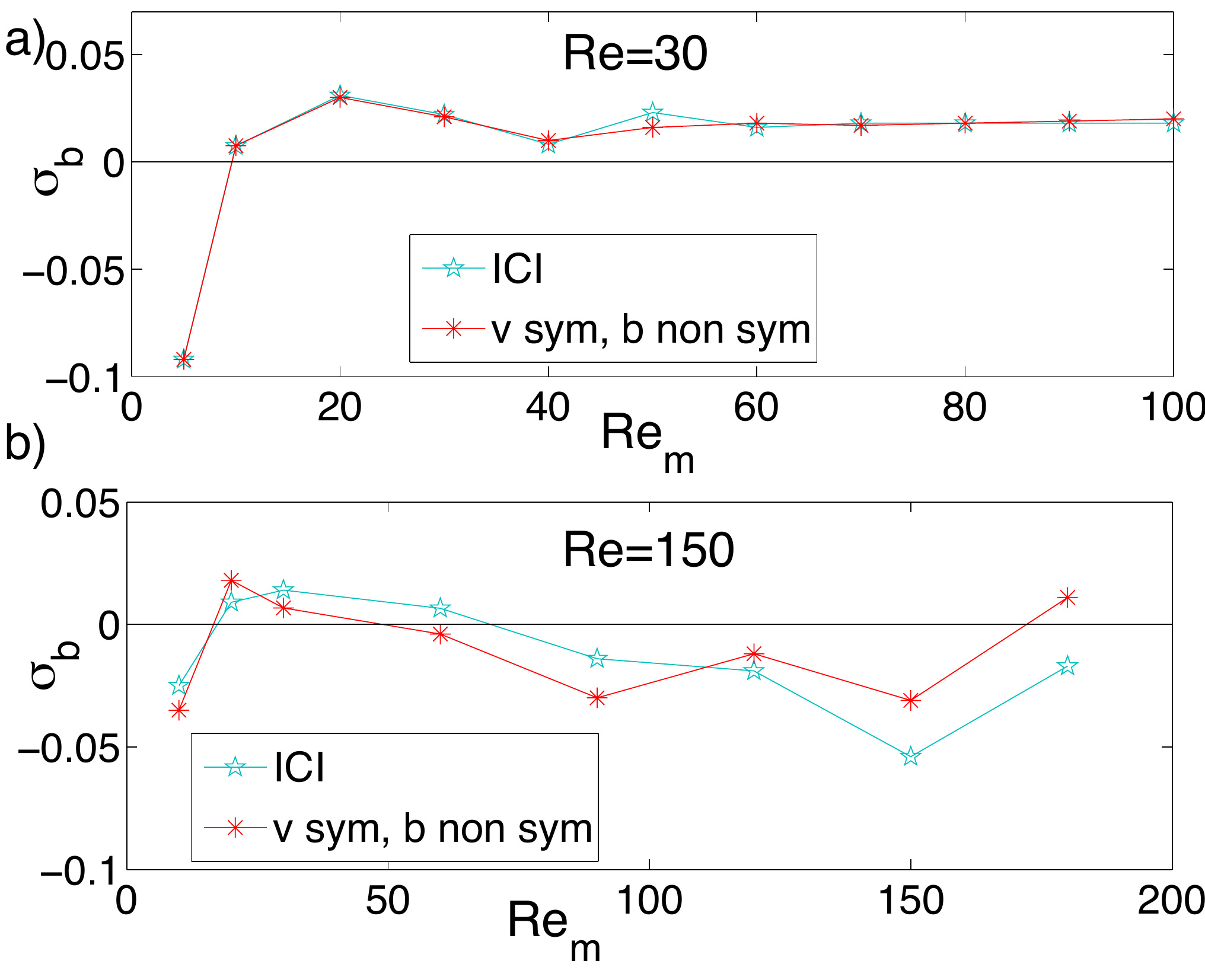}
\caption{(Color online) Dynamo growth rates $\sigma_b$ (symmetric velocity field) obtained with a non-symmetric magnetic field compared with the CIC magnetic symmetric case. Kinetic Reynolds number ${\rm Re}=30$ (a) and ${\rm Re}=150$ (b).}
\label{Fig:sigmaBsymEtNonsym}
\end{center}
\end{figure}

Note that, as a consequence of the direct sum decomposition presented in Sec.\ref{Sec:ProjTheo} (Eq.\eqref{Eq:DirectSum})), the most unstable case (ICI) and the case with no symmetries imposed on the magnetic field should have the same critical magnetic Reynolds number. This condition is indeed verified (see Fig. \ref{Fig:sigmaBsymEtNonsym}), both cases having ${\rm Re_m^{crit}}=9$. Furthermore, the property \eqref{Eq:DirectSum} also implies that the growth rates of these two cases are equal. This is apparent in Fig.\ref{Fig:sigmaBsymEtNonsym}.a, where the two growth rates are plotted for ${\rm Re}=30$. Both curves are almost identical for all magnetic Reynolds numbers used in this work. The slight difference at ${\rm Re_m}=50$ may be due to imprecisions in the numerical determination of small growth rates caused by  interferences between competing modes. At ${\rm Re}=150$ the curves only coincide qualitatively. At this relatively high kinematic Reynolds number the flow is turbulent and the instantaneous growth rate (defined as in Fig. \ref{Fig:energies}.b) fluctuates thus explaining the slight discrepancy between the two curves.

\subsection{Magnetic boundary conditions and geometry of unstable modes}

We now turn to study the geometry of the unstable modes near the dynamo threshold by generating $3D$ visualizations of the magnetic field and current.  In all the figures presented in this section the velocity field is symmetric. A resolution of $256^3$ is used and visualization are made using VAPOR. In all visualizations we represent (color online) the magnetic field lines (in red), the current (in yellow) and a density plot of the highest magnetic energy zones. The axis of forcing ($z$-axis) is indicated by a blue arrow and the equatorial directions ($x$ and $y$ axes) by red and green arrows located on one corner of the box.

\begin{figure}[h]
\begin{center}
\includegraphics[width=4.25cm,height=4.2cm]{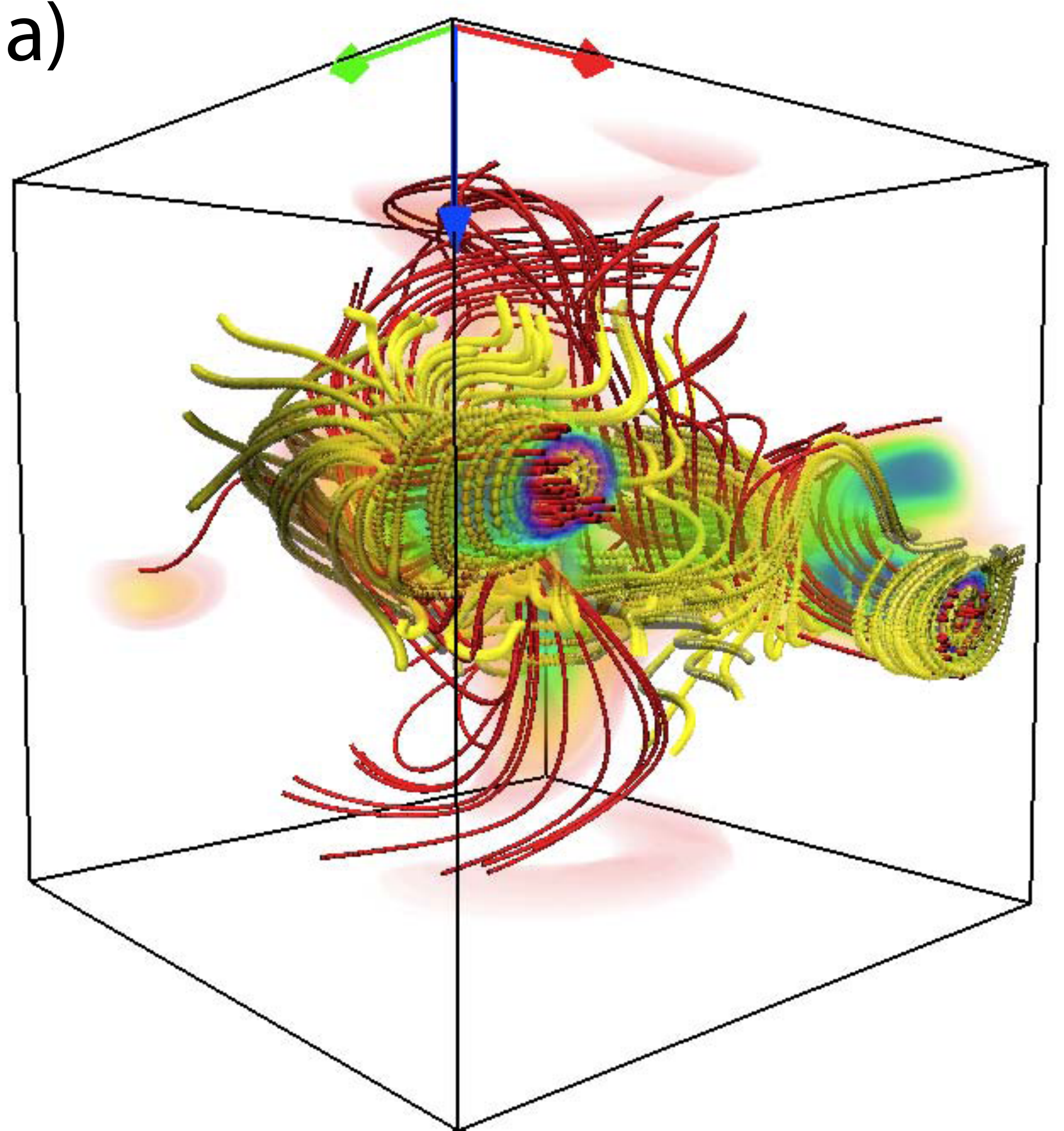}
\includegraphics[width=4.25cm,height=4.2cm]{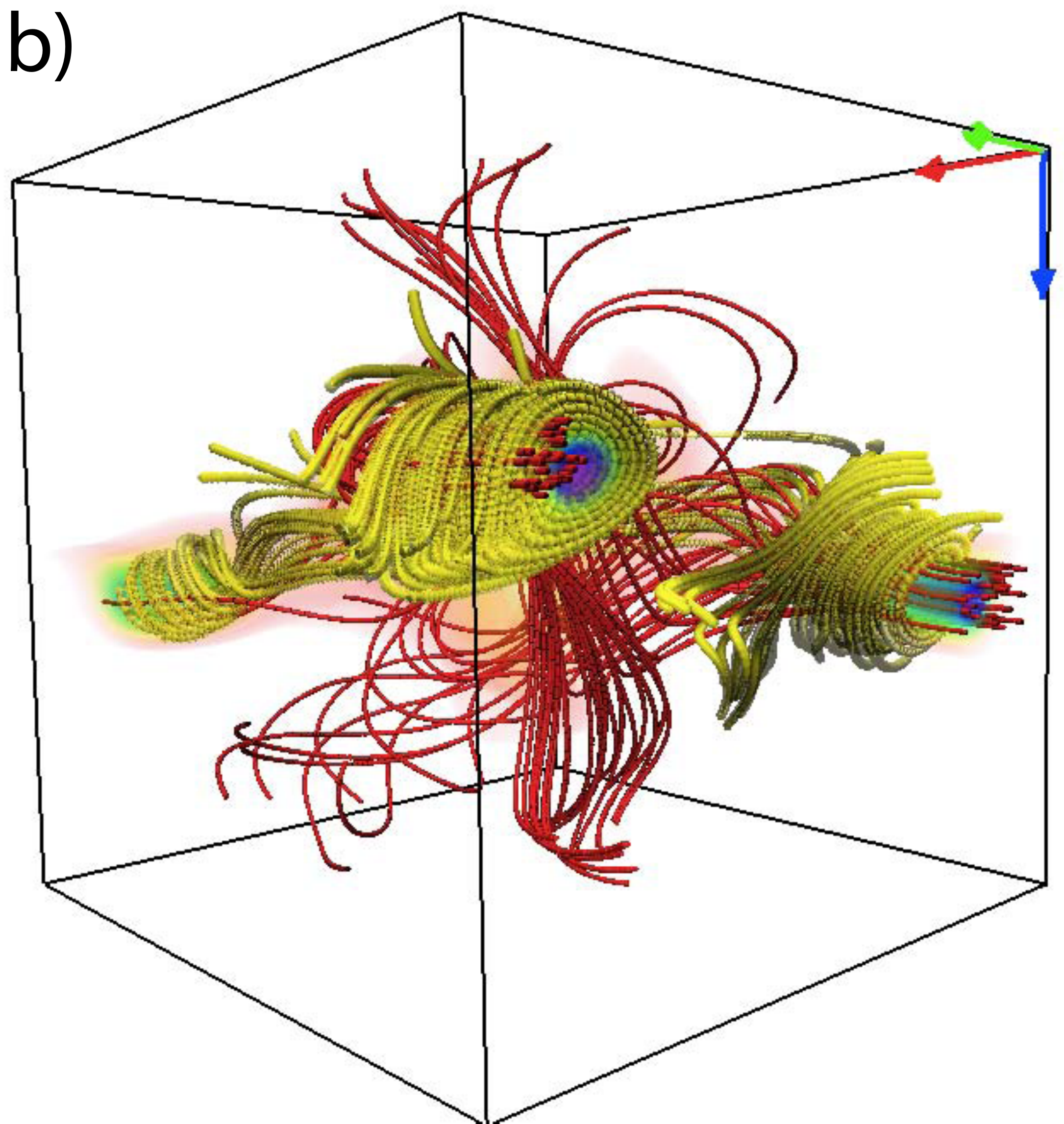}
\caption{(Color online) 3D visualizations (magnetic field in red, current in yellow and density plot of highest magnetic energy zones) of the growing modes: (a) Case $v$ symmetric and $b$ non-symmetric, ${\rm Re_m}=10$. (b) ICI case,  ${\rm Re_m}=10$.}
\label{Fig:3DvisusNonSym_ICI}
\end{center}
\end{figure}

The two cases corresponding to a non-symmetric magnetic field (a) and ICI walls (b) are first compared in Fig.\ref{Fig:3DvisusNonSym_ICI}.
Observe that, for both cases, the magnetic field is very similar: the magnetic field lines are mainly in one equatorial direction. The only difference being their respective orientation, with respect to the $x$ and $y$ axes. For the ICI case the magnetic field must cross the insulating walls, imposing its direction, while for the non-symmetric case this direction is chosen randomly by the flow between the cases ICI and CIC (recall that the cases ICI and CIC are equivalent by  $\pi/2$ rotation). This field is very similar to the one reported in \cite{Nore:1997p5767}.

\begin{figure}[h]
\begin{center}
\includegraphics[width=4.25cm,height=4.25cm]{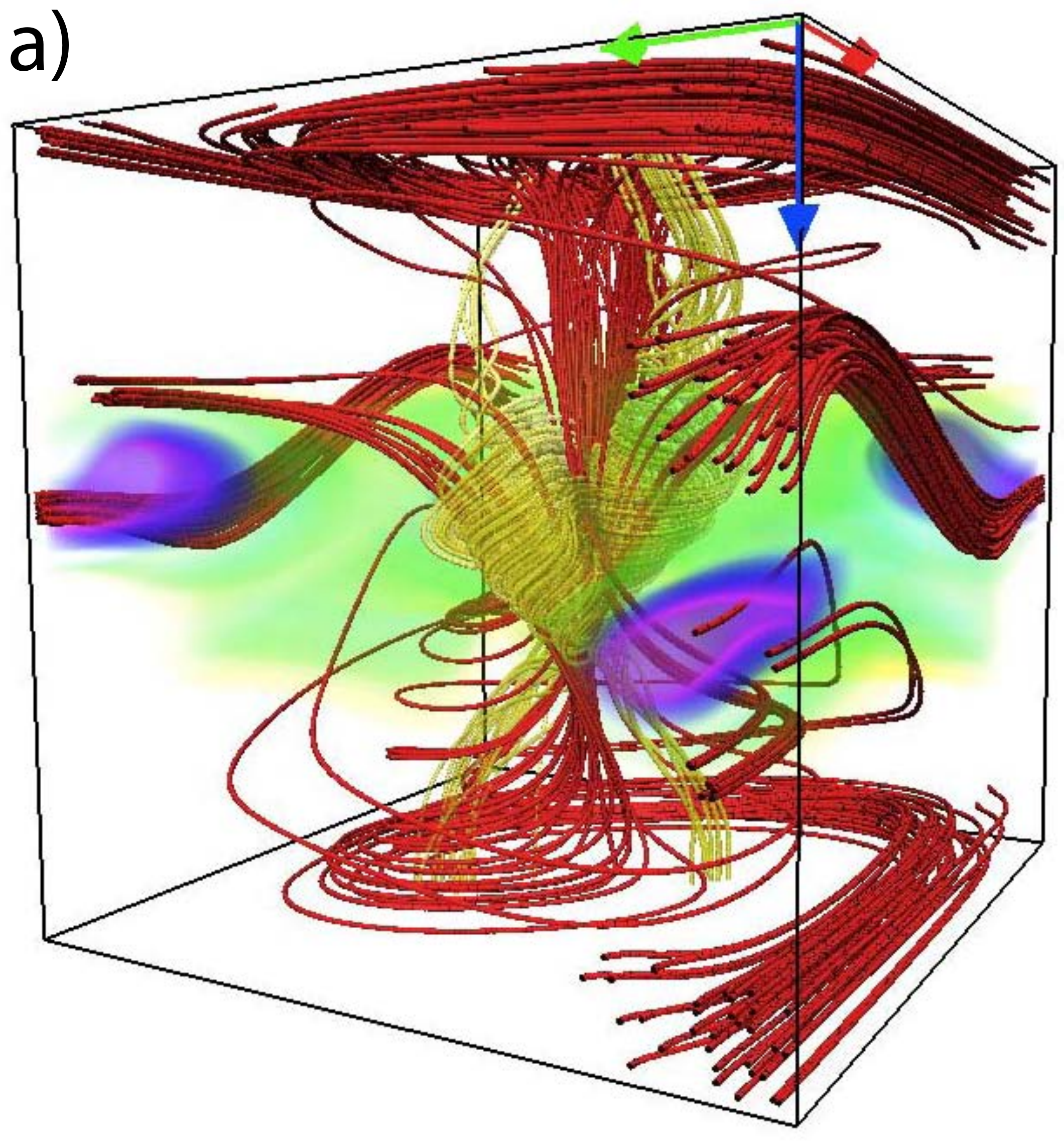}
\includegraphics[width=4.25cm,height=4.25cm]{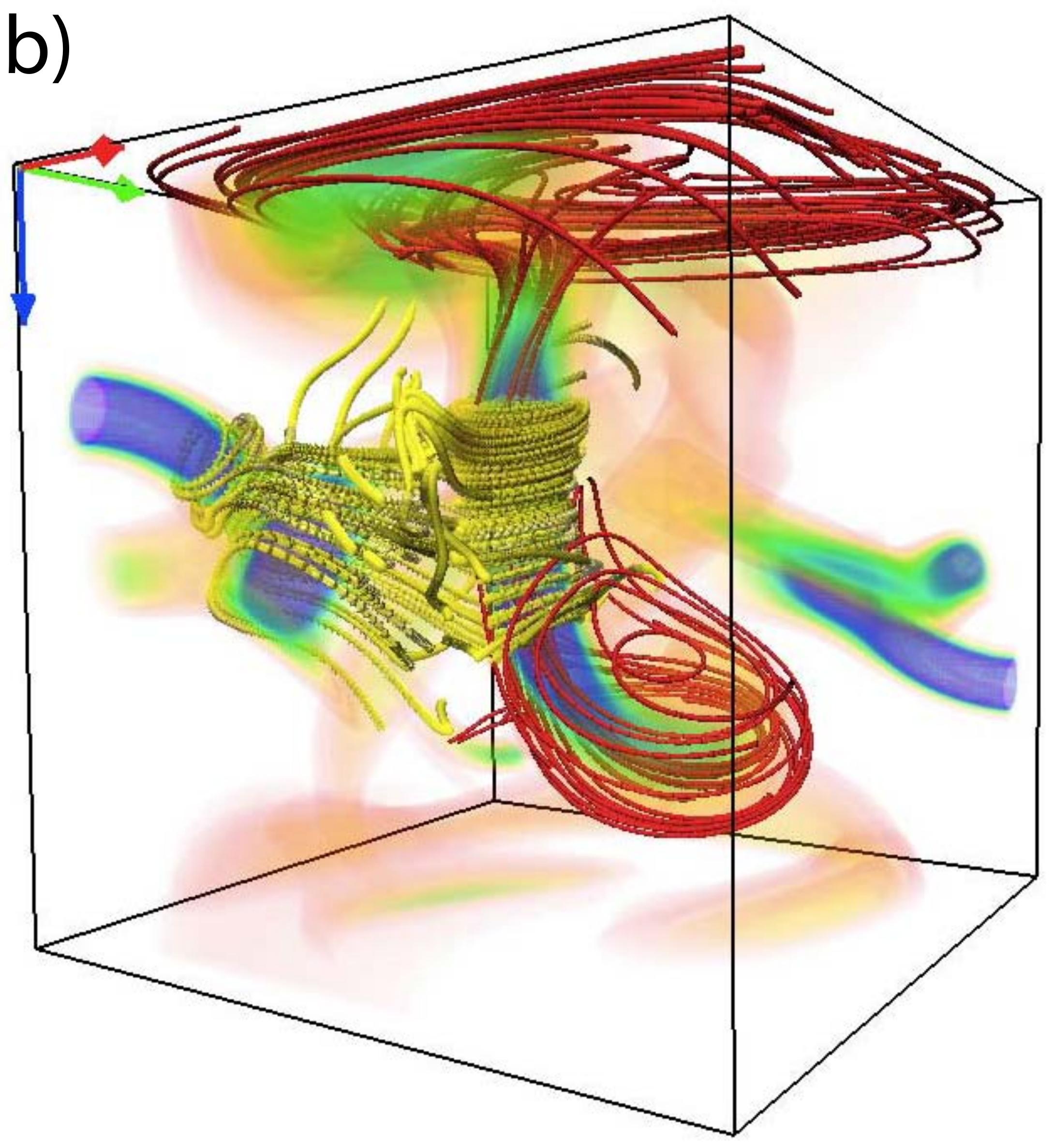}
\includegraphics[width=4.25cm,height=4.25cm]{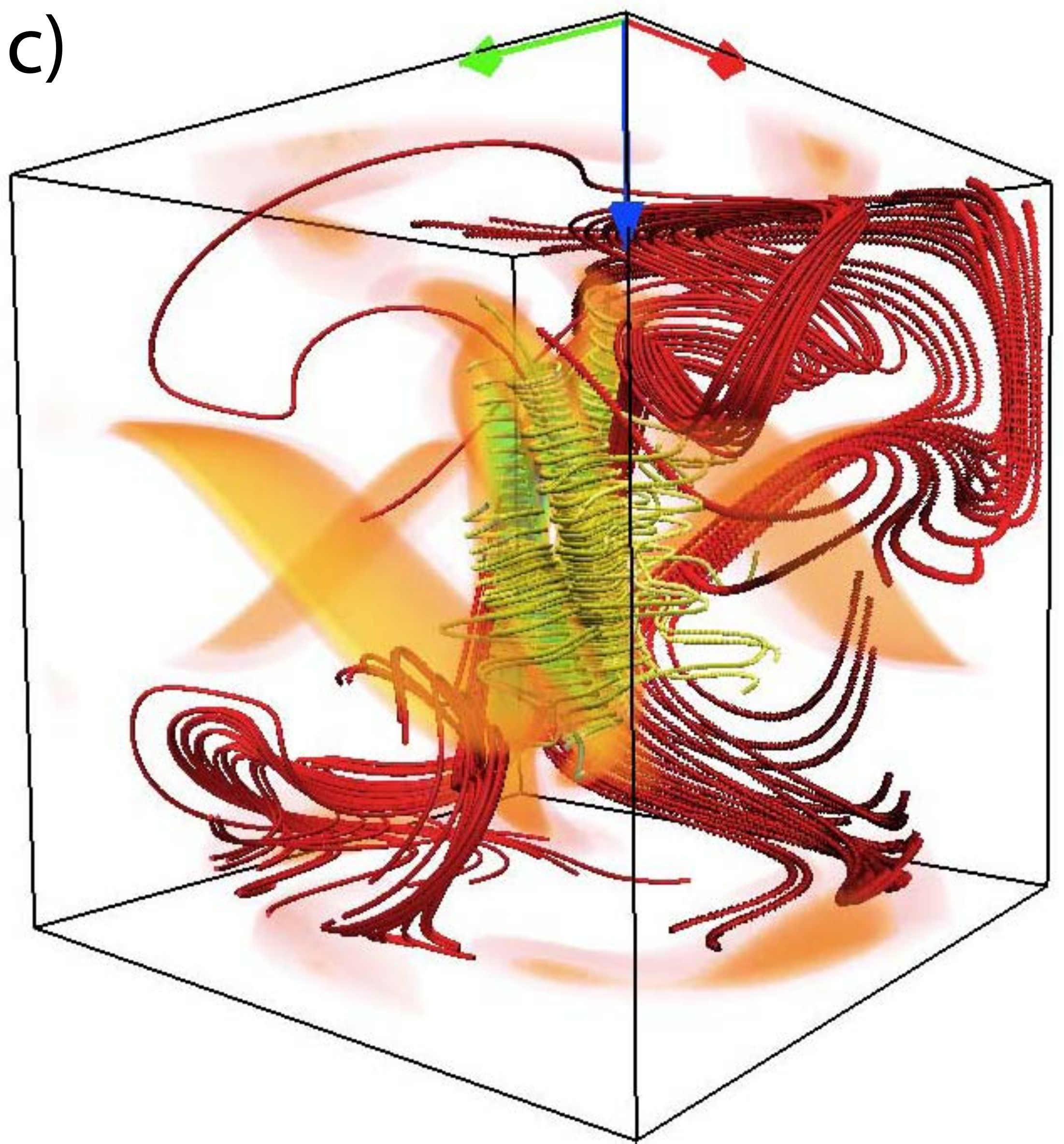}
\includegraphics[width=4.25cm,height=4.25cm]{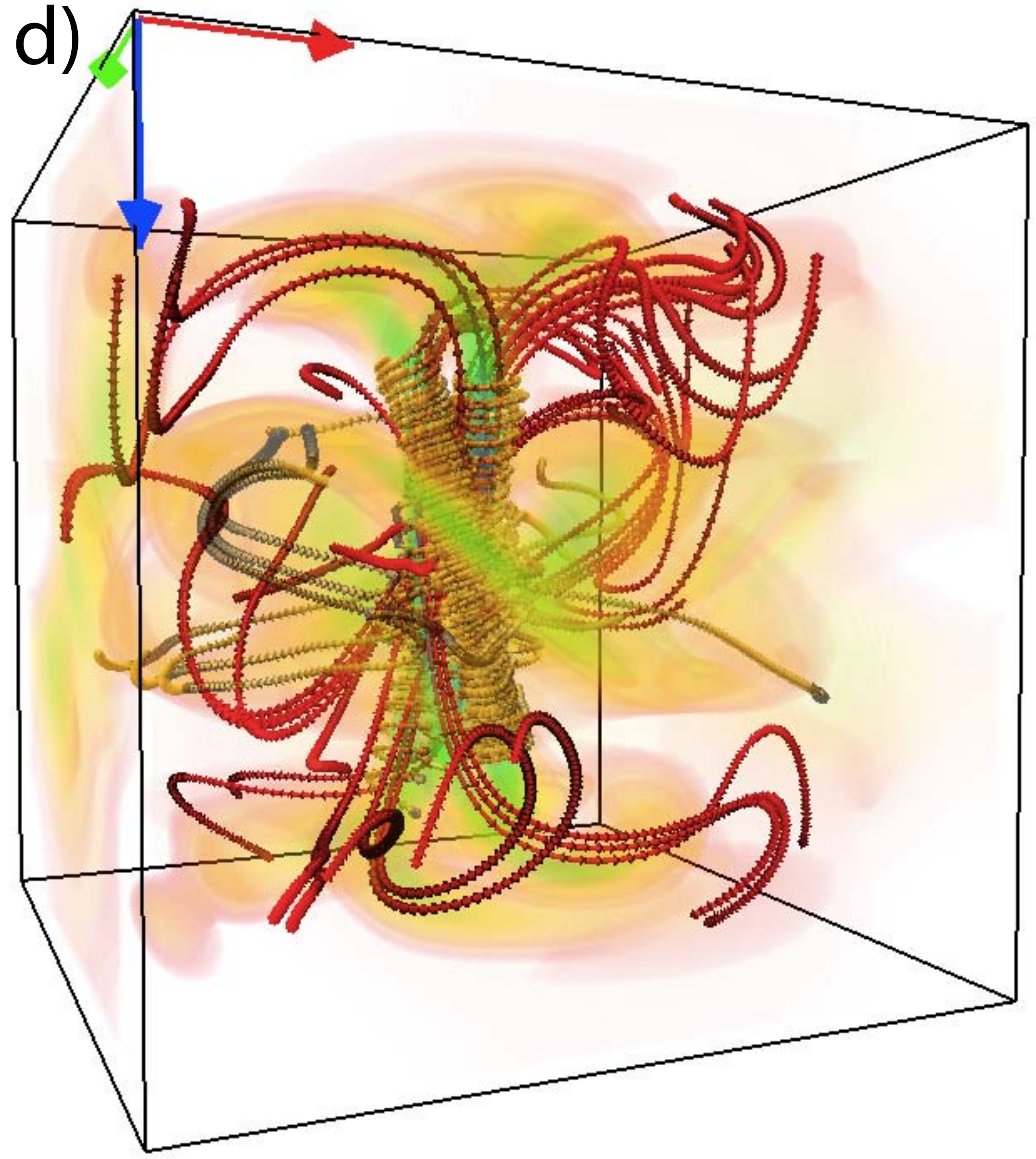}
\caption{(Color online) 3D visualizations (magnetic field in red, current in yellow and density plot of highest magnetic energy zones) of the growing modes: (a) ICC case,  ${\rm Re_m}=30$ (b) IIC case,  ${\rm Re_m}=80$, (c) CCC case,  ${\rm Re_m}=300$, (d) CCI case,  ${\rm Re_m}=300$.
}
\label{Fig:3Dvisus4cases}
\end{center}
\end{figure}

The visualization of cases ICC, IIC, CCC and CCI are displayed in Fig.\ref{Fig:3Dvisus4cases}.
Note that both ICC and IIC cases contain axial magnetic fields but the magnetic field lines do not cross the top and the bottom of the box, a possibility that is not allowed by the boundary conditions (C walls on top and bottom). The cases CCC and CCI present complicated structures.

\begin{figure}[h]
\begin{center}
\includegraphics[width=4.25cm]{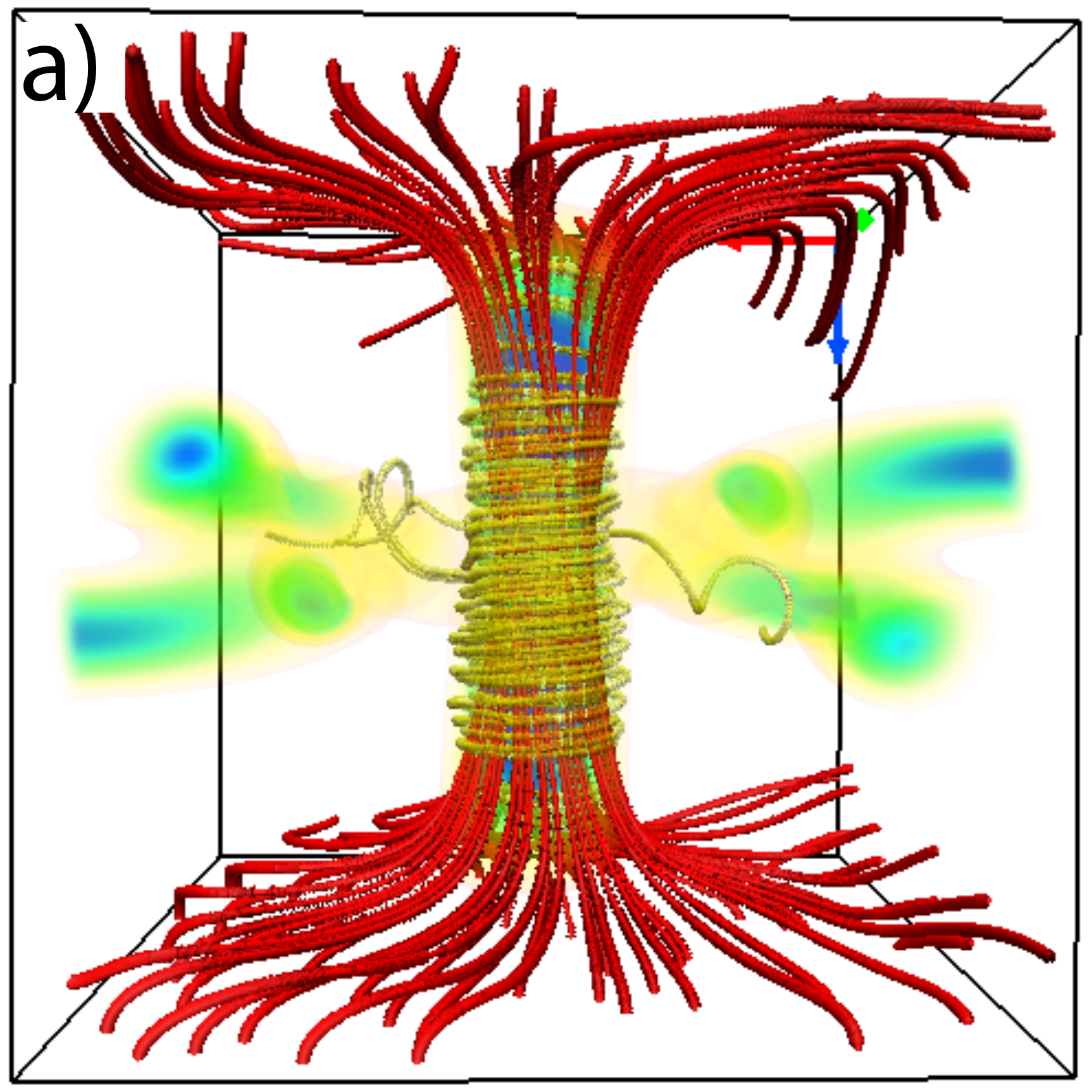}
\includegraphics[width=4cm]{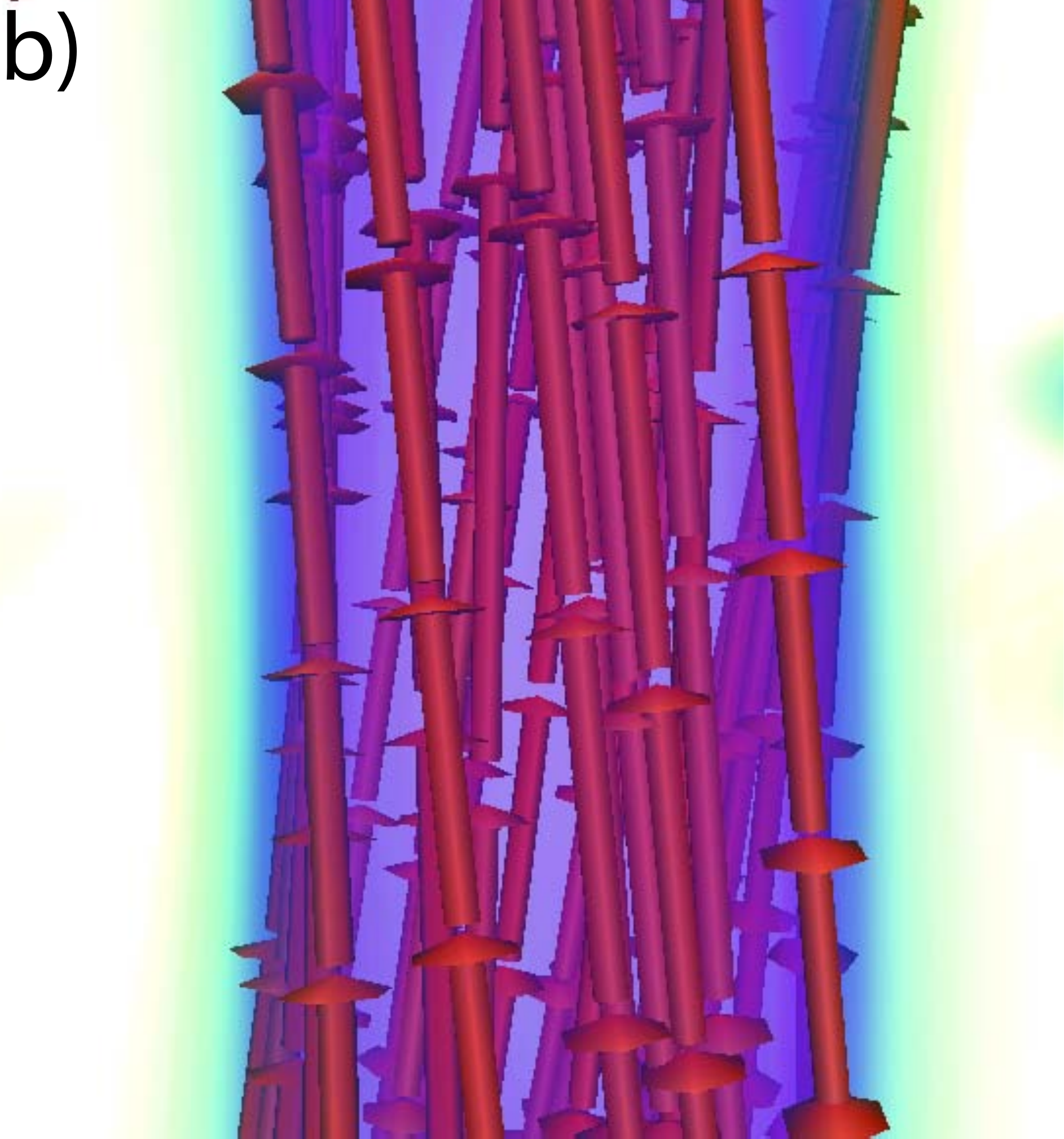}
\caption{(Color online) 3D visualizations (magnetic field in red, current in yellow and density plot of highest magnetic energy zones) of the growing modes: (a) III case,  ${\rm Re_m}=80$. (b) Zoom of figure (a) at the center of the box.}
\label{Fig:3DvisusIIIcase}
\end{center}
\end{figure}

Fig. \ref{Fig:3DvisusIIIcase}.a displays another of the main results of this paper: an axial dipole that is obtained with III walls. The magnetic lines are clearly oriented along the $z$-axis (see the zoom view displayed on  \ref{Fig:3DvisusIIIcase}.b). 
To the best of our knowledge this is the first time that an axial dipole has been observed using periodical boundary conditions.

\section{Non-linear saturation of axial dipolar magnetic field}\label{sec:saturation}

We have studied so far the kinematic dynamo problem and have found how the first unstable magnetic mode, when ${\rm Re_m}$ is increased, depends on the boundary conditions. We present in this section the different stationary regimes that are obtained when the growing unstable mode saturates due to the back reaction of the Lorentz force on the velocity field. This study is restricted to the case III for which an axial dipole is the first unstable magnetic mode.

\begin{figure}[h]
\begin{center}
\includegraphics[width=8.25cm]{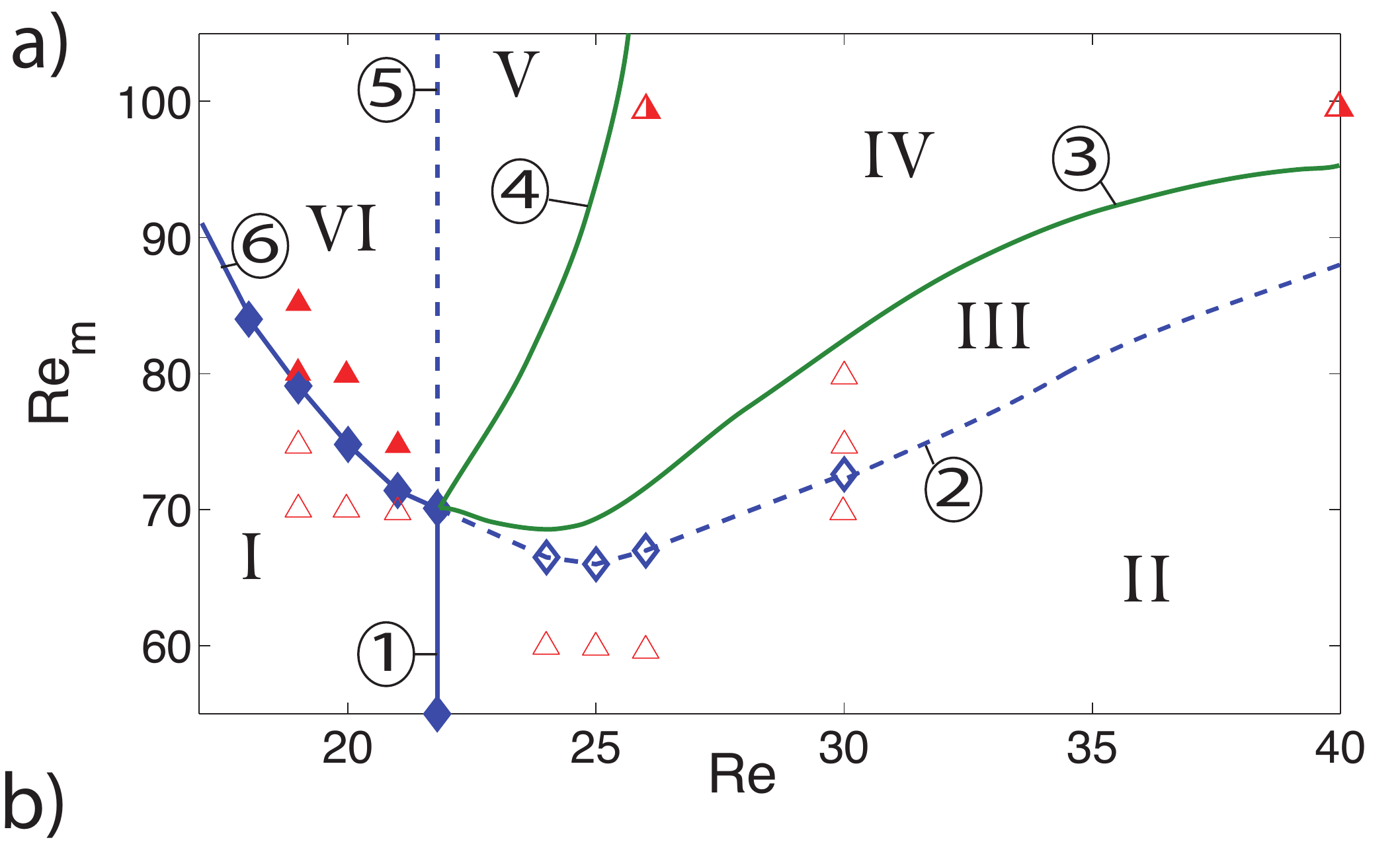}
\includegraphics[width=8.25cm]{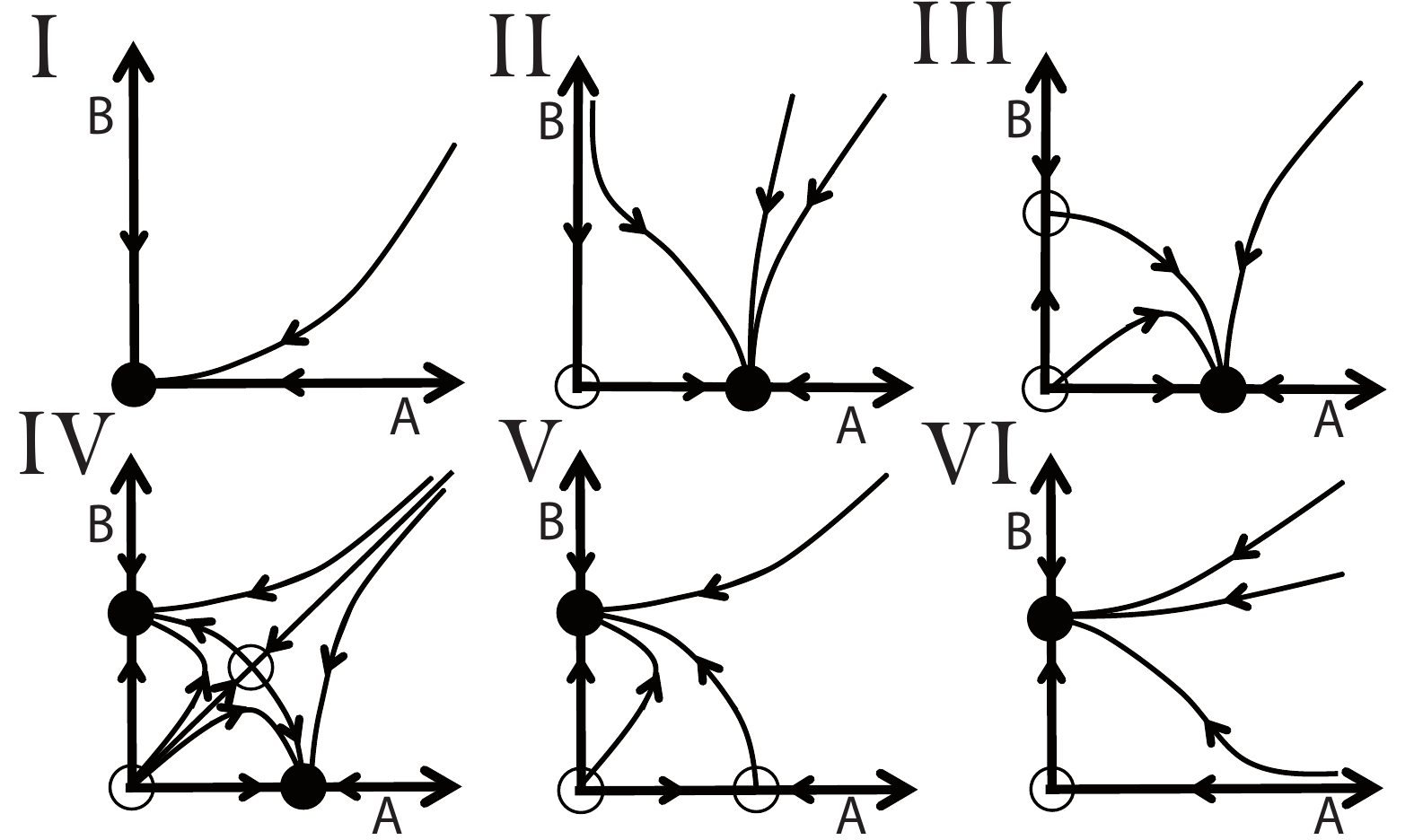}
\caption{(Color online) a) Qualitative drawing of the bifurcation lines. In the vicinity of the codimension-two point where these lines intersect, they are given by model \eqref{EQ:systdyn}. Diamonds correspond to data coming from numerical simulations (blue lines), lines without markers are qualitatively drawn (green lines).  Each bifurcation line is labeled with a number (see text) and each region of parameter space is labeled by I-VI . Red triangles corresponds to some simulations used to identify the different regions of the parameter space. Filled (empty) markers denote a non-vanisnhing (vanishing) magnetic field in saturated regimes. Half-filled triangles denote points in the bi-stable zone. b) Phase diagrams (in $A$-$B$ space, see section \ref{sec:codim2}) corresponding to regions I-VI of (a).}
\label{Fig:Model}
\end{center}
\end{figure}

The critical ${\rm Re_m^{crit}}$ bifurcation line, observed in numerical dynamo simulations (III case) around the stable branch of section \ref{sec:symbreakpis2} (see Fig. \ref{Fig:TGvis}.a), is presented in Fig. \ref{Fig:Model}.a (lower curve ($6$ and $2$) with diamonds). The vertical line at  ${\rm Re^c}=22$  ($1$ and $5$) corresponds to the hydrodynamic pitchfork bifurcation discussed in section \ref{sec:symbreakpis2}.
These two bifurcation lines intersect at a so-called codimension-$2$ bifurcation point. Dynamical regimes in the vicinity 
of this point are displayed as the phase portraits of Fig.\ref{Fig:Model}.b that will be discussed below.

\subsection{Dynamo shutdown by velocity bifurcation}

\begin{figure}[h]
\begin{center}
\includegraphics[width=8.4cm]{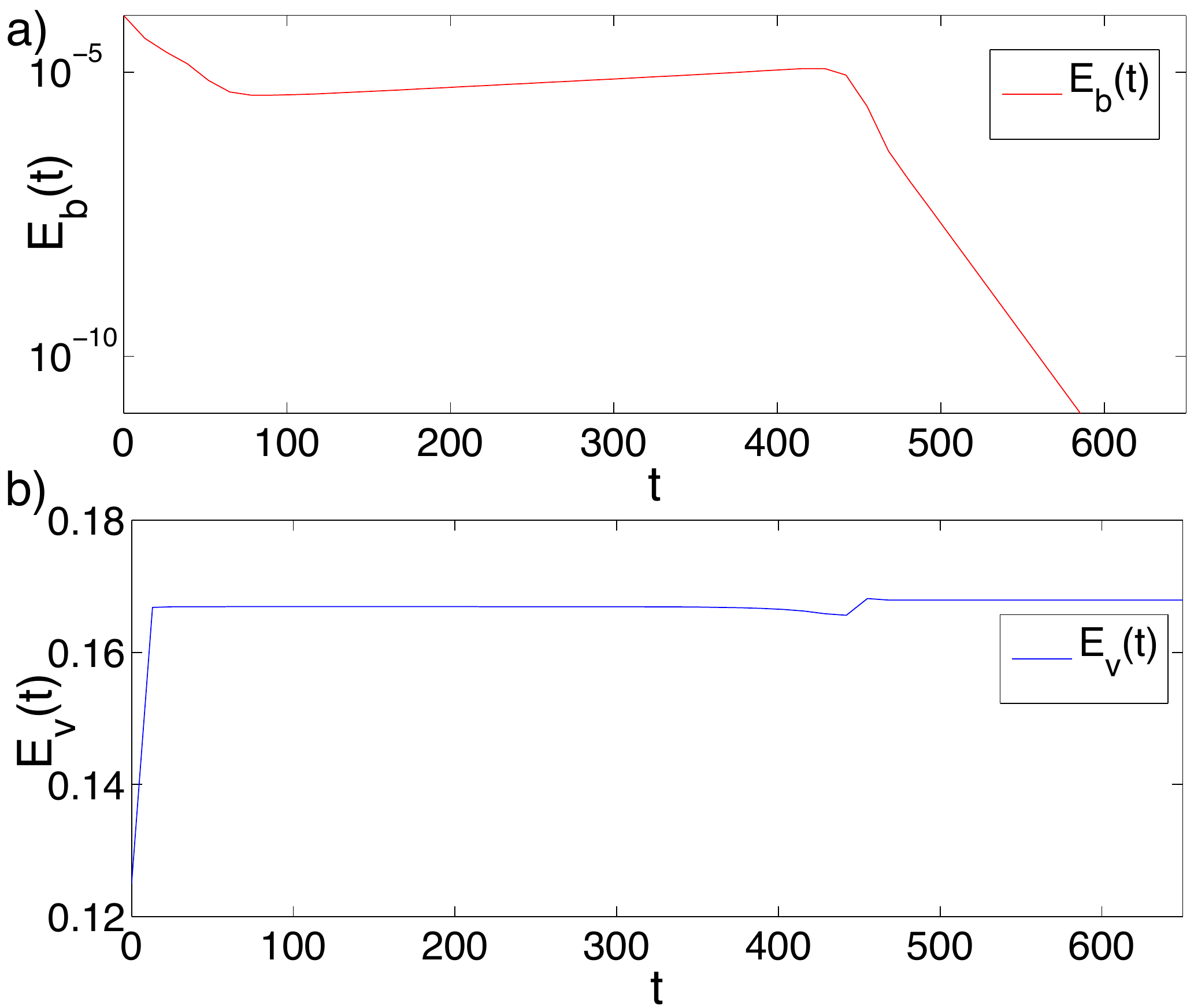}
\caption{Case III: Temporal evolution of kinetic and magnetic energy at ${\rm Re}=30$ and ${\rm Re_m}=80$. Vanishing of saturated magnetic field is clearly observed at large times. }
\label{Fig:effMag}
\end{center}
\end{figure}

Let us first emphasize that the nonlinearly saturated regime is not always related to the linearly growing mode. This type of behavior, obtained for ${\rm Re}=30$, is displayed in Fig.\ref{Fig:effMag}.a. For ${\rm Re_m}=80$, {\it i.e.} above the bifurcation threshold ${\rm Re_m^{\rm{crit}}}=73$ (region III in Fig.\ref{Fig:Model}.a), the magnetic energy displays a growing phase (after some transient) but then  decreases and vanishes in the long time limit. Figure \ref{Fig:effMag}.b shows that this change of behavior of the magnetic energy occurs concomitantly with a variation of the kinetic energy of the flow. 

We have checked that this variation is related to the transition from the symmetric flow taken as initial conditions to the one that breaks the $\pi/2$ rotational invariance around the axis $(x=y=\pi/2)$ (see section \ref{sec:symbreakpis2}). Although the symmetric flow is hydrodynamically unstable for ${\rm Re}=30$, the initial conditions are such that it has not yet broken its symmetry when the magnetic field begins to grow. It is the Lorentz force that drives the flow to one of its bifurcated states with broken $\pi/2$ rotational invariance. The bifurcated flow having no dynamo capability for ${\rm Re_m}=80$, the magnetic field then decays. Note that a similar phenomenon was observed for the magnetic field generated by a flow in a spherical domain \cite{fuchs1999}.

\subsection{Supercritical dynamo}

\begin{figure}[h]
\begin{center}
\includegraphics[width=8.6cm]{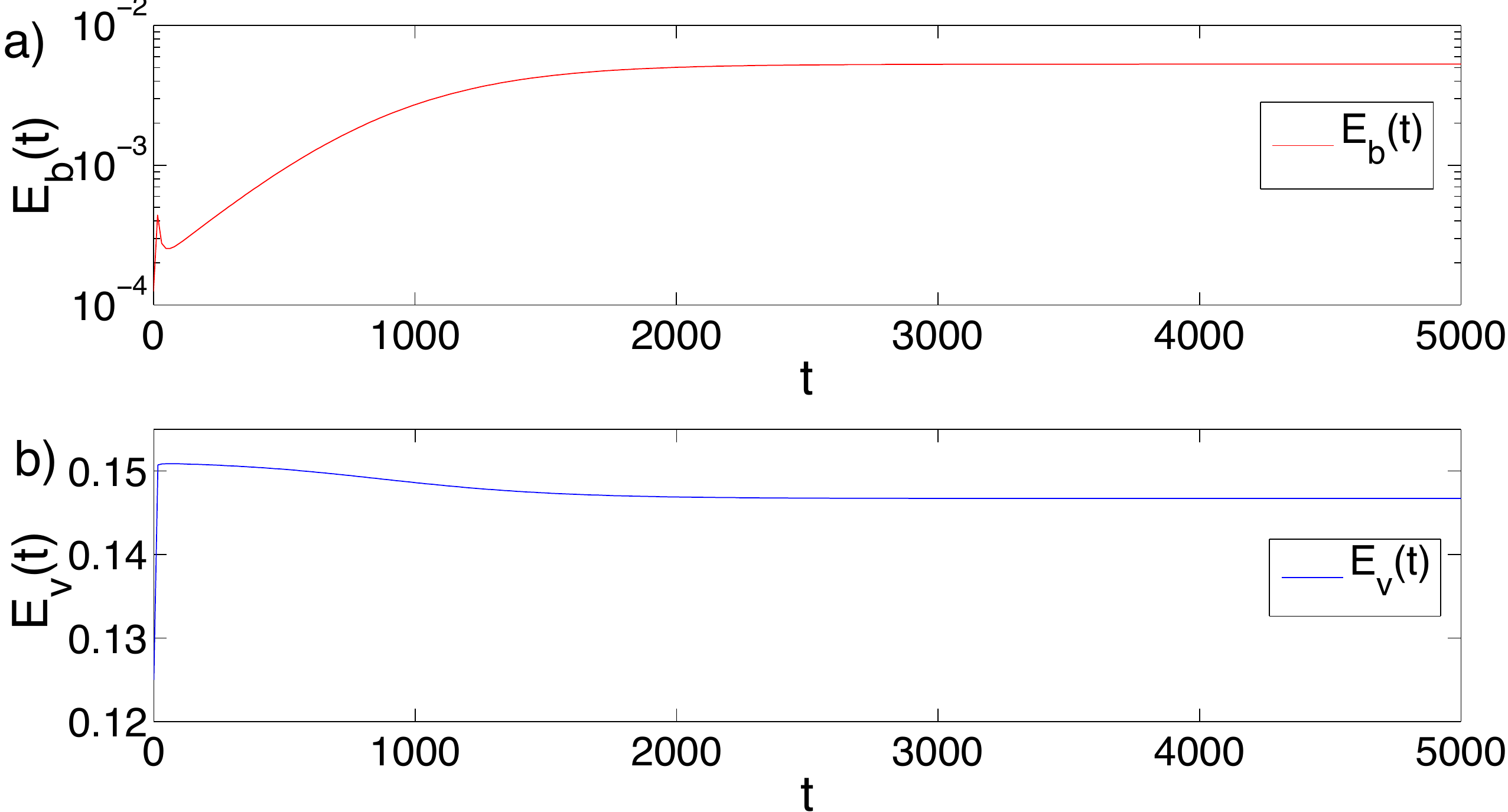}
\includegraphics[width=4.cm]{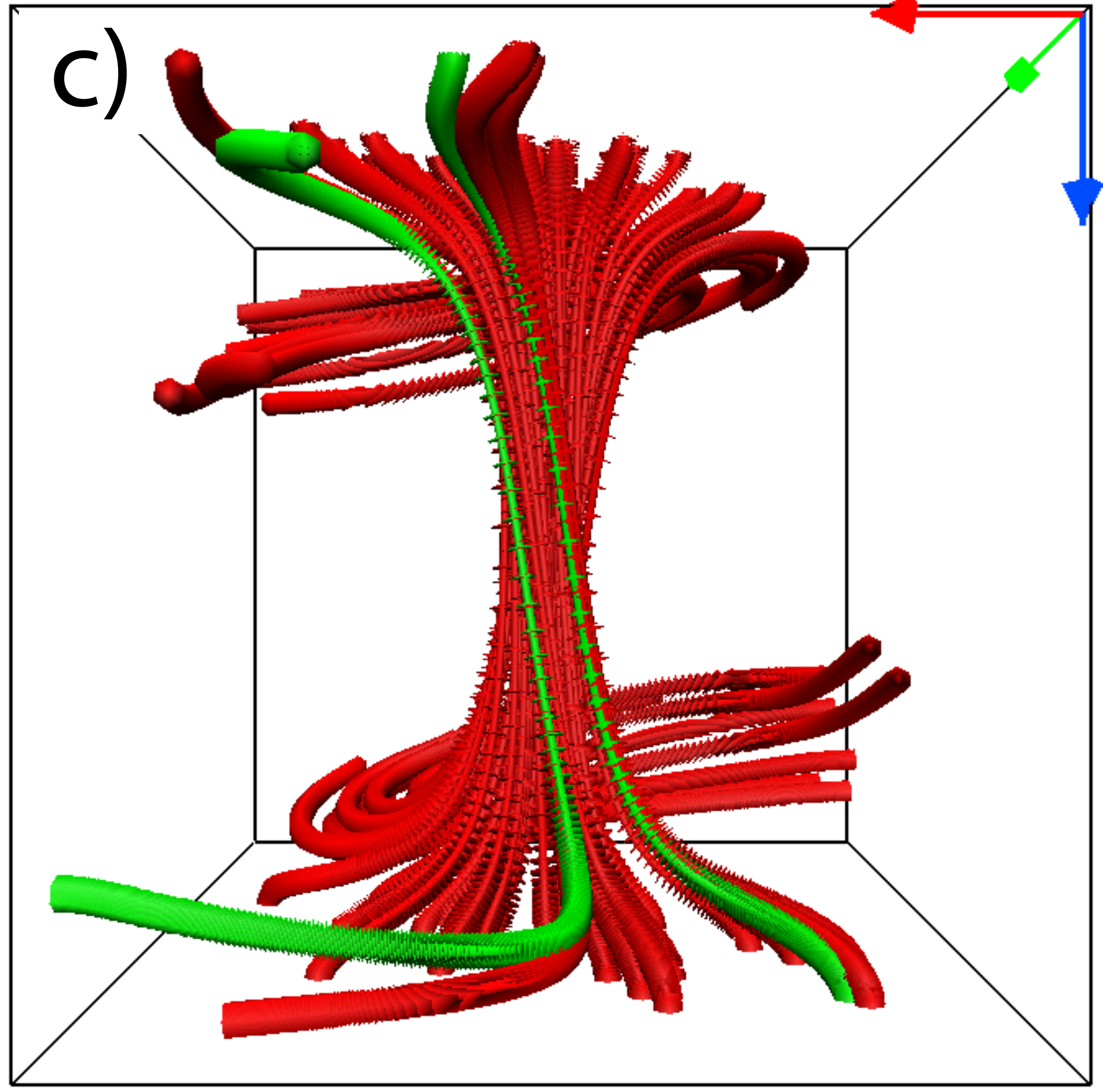}
\includegraphics[width=4.cm]{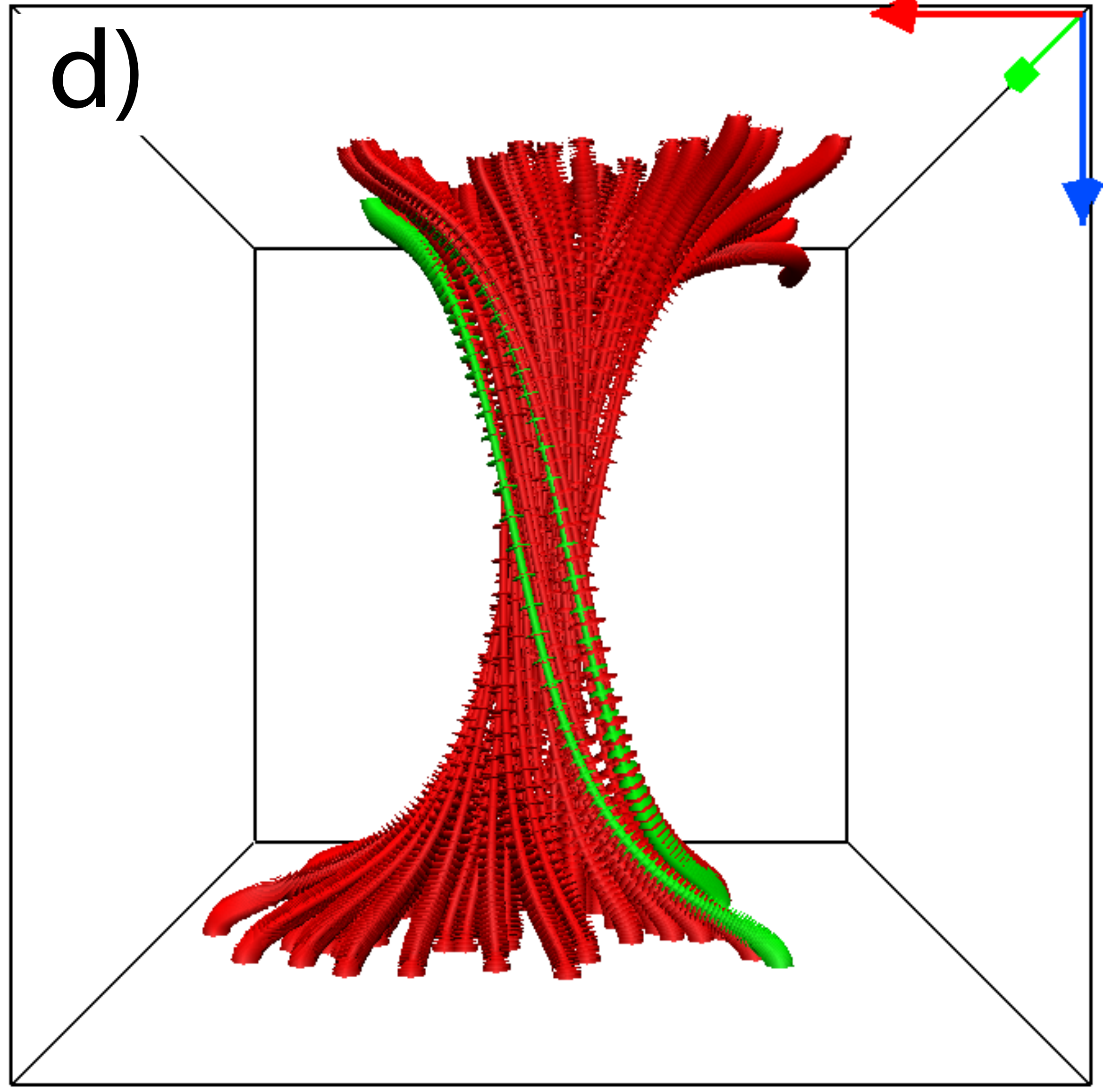}
\caption{(Color online) Case III:  Temporal evolution of magnetic  (a) and kinetic energy (b) at ${\rm Re}=20$ and ${\rm Re_m}=80$. c) $3D$ visualization of magnetic field at linear growth phase $t=400$. d) Visualization  of saturated magnetic field at  $t=3000$. Two of the magnetic field lines have been colored in green in order to emphasize the twisting of the saturated field.}
\label{Fig:effMag2}
\end{center}
\end{figure}

The complex transient behavior of the magnetic field being related to the interaction with the hydrodynamic pitchfork bifurcation reported in section \ref{sec:symbreakpis2}, we now study the dynamo bifurcation at ${\rm Re}=20$, below this hydrodynamic bifurcation threshold (region VI in Fig.\ref{Fig:Model}.a). 
As shown in Fig. \ref{Fig:effMag2}.a the linear unstable magnetic mode saturates at finite amplitude thus displaying a supercritical pitchfork bifurcation. 
The kinetic energy on Fig. \ref{Fig:effMag2}.b is slightly reduced by the saturation mechanism.
It is also apparent on Fig. \ref{Fig:effMag2}.c-d that the magnetic field lines are slightly more twisted in the saturated regime, as can be checked visually by focusing on the two magnetic field lines that are colored in green in the figure.

We now consider how the magnetic energy \eqref{eq:EnergiesDef} depends on the fluid parameters above the dynamo threshold. Dimensional arguments imply $\langle {\bf b}^2 \rangle =  \langle {\bf v}^2 \rangle \, f({\rm Re}, {\rm Re_m})$ where $f$ is an unknown function. Close to a supercritical bifurcation threshold, we expect that $\langle {\bf b}^2 \rangle$ depends linearly on ${\rm Re_m} - {\rm Re_m^{\rm{crit}}}$, thus
\begin{equation}
\begin{aligned}
\langle {\bf b}^2 \rangle \simeq  \langle {\bf v_c}^2 \rangle \, g({\rm Re}) \,  \frac{\rm Re_{m}- \rm Re_{m}^{crit}}{\rm Re_{m}^{crit}}, \label{EQ:scalingb} 
\end{aligned}
\end{equation}
where $\langle {\bf v_c}^2 \rangle$ is related to the kinetic energy density at bifurcation threshold. It is expected that $g({\rm Re})$ tends to a constant in the limit of large ${\rm Re}$ and is inversely proportional to ${\rm Re}$ for small ${\rm Re}$ \cite{petrelis2001}. Using $v_{\rm rms}^2=2 E_v/3$ to estimate $\langle {\bf v_c}^2\rangle$, we find that $g({\rm Re}) \sim 3$ 
increases by about $20 \%$ when ${\rm Re}$ is varied from $18$ to $21$ (data not shown). 
The high ${\rm Re}$ number scaling thus is observed even at moderate values of ${\rm Re}$ (compared to experiments with liquid metals).

\subsection{Bistability}

For ${\rm Re}>22$, a bistable region can be found in which, depending on the initial conditions, we can get both a dynamo and purely hydrodynamic regimes.
For instance, this case was observed (data not shown) at ${\rm Re}=26$ and ${\rm Re_m}=100$ (region IV in Fig.\ref{Fig:Model}.a).

Starting from a dynamo in such a bistable regime, we followed a line at ${\rm Re_m}=100$ by increasing the Reynolds number up to ${\rm Re}=120$ without loosing the dynamo (data not shown).

A hysteresis was also observed, for ${\rm Re}>22$, by varying ${\rm Re_m}$ at fixed ${\rm Re}$ (data not shown).

\subsection{Codimension-$2$ bifurcation model} \label{sec:codim2}

We next present a simple explanation for the super/sub critical nature of the dynamo transition as the kinetic Reynolds number (or equivalently the magnetic Prandtl number of the fluid) is varied. In our simulations, this phenomenon is strongly related to the presence of a pitchfork bifurcation of the flow for ${\rm Re} = {\rm Re^c} = 22$. 

For ${\rm Re} = {\rm Re^c}$ and ${\rm Re_m} ={\rm Re_{m}^{crit}} ({\rm Re_c}) ={\rm  Re_m^c}$, we have a codimension-two bifurcation. In its vicinity, the dynamo and the hydrodynamic instabilities compete, thus generating various dynamical regimes. 

Denoting by $A(t)$ the real amplitude of the bifurcating velocity field ${\bf v^{\rm PF}}$ and $B(t)$ the real amplitude of the bifurcating magnetic field $\bf b$, we write coupled amplitude equations for $A$ and $B$ in the vicinity of the codimension-two bifurcation. The form of these equations is constrained by symmetry requirements, $A \rightarrow -A$ (pitchfork bifurcation of the velocity field) and $B \rightarrow -B$ ($\bf b \rightarrow - \bf b$ symmetry of the MHD equations).  

Keeping the nonlinear terms to leading order, we get
\begin{eqnarray}
\dot A&=&\lambda A - \alpha A B^2 - A^3,\nonumber \\
\dot B&=&\mu B - \beta A^2 B - B^3.  \label{EQ:systdyn}
\end{eqnarray}
The coefficients of the cubic nonlinearities have been taken negative in order to get supercritical pitchfork bifurcations for the hydrodynamic instability in the absence of magnetic field ($B=0$) and for the dynamo instability when $Re < Re^c$ and thus $A=0$. The modulus of these coefficient can be taken equal to $1$ by appropriate scalings of the amplitudes $A$ and $B$. $\lambda$ and $\mu$ are functions of $Re$ and $Re_m$ that vanishes at the codimension-two bifurcation point $(Re^c, Re_m^c)$. To leading order, we have $\lambda \propto Re - Re^c$ and $\mu \propto Re_m - Re_m^{crit}$.  

The fixed points of the system \eqref{EQ:systdyn} are $(0, 0)$, $(\pm \sqrt \lambda, 0)$, $(0, \pm \sqrt \mu)$ and the mixed modes $(\pm \sqrt{(\lambda - \alpha \mu)/(1 - \alpha \beta)}, \pm \sqrt{(\mu - \beta \lambda)/(1 - \alpha \beta)}$. The different types of bifurcation diagrams  have been studied in detail \cite{guckenheimer1983}. 

The cases of interest are presented in Fig.\ref{Fig:Model}.b. They correspond to $\alpha \beta > 1$ (in order to have unstable mixed modes) with $\alpha$ and $\beta$ positive (in order to prevent the existence of subcritical mixed modes for $\lambda$ and $\mu$ negative). Then the globally stable solution $(0, 0)$ for $\lambda < 0$ and $\mu < 0$ undergoes a supercritical pitchfork bifurcation to $(\pm \sqrt \lambda, 0)$ (respectively $(0, \pm \sqrt \mu)$) for $\lambda = 0$ (respectively $\mu = 0$). 

The corresponding bifurcation lines are labelled $(1)$ (respectively $(6)$ in Fig. \ref{Fig:Model}). For $\lambda > 0$ (respectively $\mu > 0$), the second pure mode bifurcates for $\mu = 0$ (respectively $\lambda = 0$) but is unstable (bifurcation lines $(2)$ and $(5)$). 

The mixed modes exist between the lines $\lambda = \alpha \mu$ and $\mu = \beta \lambda$ (corresponding to $(3)$ and $(4)$ in Fig. \ref{Fig:Model}). The key point is that they are unstable with respect to the pure modes. The system is thus bistable in this parameter range. 

The bifurcated hydrodynamic regime as well as the dynamo state are both linearly stable. When ${\rm Re_m}$ is increased for ${\rm Re} > {\rm Re^c}$, the hydrodynamic regime bifurcates to a dynamo state on line $(4)$. If ${\rm Re_m} $ is then decreased from this state, the dynamo is suppressed on line $(3)$, thus displaying an hysteresis. 

An hysteresis can be also observed by varying ${\rm Re}$ for ${\rm Re_m }> {\rm Re_m^c}$. If the fluid velocity is increased at constant magnetic Prandtl number, i.e. if one follow a line ${\rm Re_m }={\rm  P_m Re}$ in parameter space, we expect a supercritical dynamo bifurcation for ${\rm P_m}$ large enough and a subcritical one for ${\rm P_m}$ small.

Let us finally remark that the model \eqref{EQ:systdyn} is expected to be valid only in a neighborhood of the codimension-two point  $(Re^c, Re_m^c)$. Indeed many other secondary bifurcations can take place away from this point.  Because of this and also due to the limited number of runs that were performed, the lines in Fig.\ref{Fig:Model}.a are only qualitatively drawn.

\section{Discussion and conclusion}\label{sec:conclusion}

It has been often claimed that too many symmetries of the velocity and/or magnetic fields inhibit dynamo action. This claim probably results from several anti-dynamo theorems  that have been found since Cowling who showed that an axisymmetric velocity field cannot generate an axisymmetric magnetic field (for a review, see \cite{nunez1996}). Thus, the magnetic field should break axisymmetry when generated through dynamo action by an axisymmetric velocity field. Another class of anti-dynamo theorems is even more restrictive and forbids the dynamo action of some velocity fields (for instance planar flows, {\it i.e.} velocity fields with only two non zero cartesian components) whatever the geometry of the generated magnetic field. 

We have shown here that symmetries do not always inhibit dynamo action, but, on the contrary can sometimes enhance it: 

- symmetry constraints on the velocity field can lead to lower dynamo threshold by inhibiting the development of hydrodynamic instabilities and related turbulent fluctuations that sometimes reduce the efficiency of dynamo action,

- symmetry constraints on the magnetic field can lead to an unchanged dynamo threshold provided they are chosen in the appropriate manner. 

It has been shown that the dynamo threshold ${\rm Re_m^{crit}}$ of the TG flow, increases when the kinetic Reynolds number of the flow is increased on some intermediate range, and this has been related to the development of turbulent fluctuations \cite{ponty2005,laval2006,ponty2007}.   Using symmetry constraints on the velocity field shows that phenomenon without ambiguity: the level of turbulent fluctuations is lower with the symmetric velocity field and, correspondingly, the dynamo threshold is lower too, although the mean kinetic energy is larger for the velocity field without symmetries. This shows in a simple and clear-cut way that velocity fluctuations inhibit dynamo action by the TG flow. The same phenomenon has been shown analytically for the dynamo generated by a fluctuating Roberts flow \cite{petrelis2006}. However, it should be kept in mind that a time-dependent velocity can also generate new dynamo modes that do not exist in the absence of fluctuations.

We have also shown that symmetry constraints on the magnetic field can be used to mimic realistic boundary conditions in the framework of numerical simulations with periodic codes.
It was found that the dynamo threshold and the geometry of the growing magnetic mode strongly depend on the choice of symmetry constraints, and thus on the related boundary conditions. 
The lowest dynamo threshold was obtained with lateral boundaries of different nature, allowing a magnetic field to cross the box perpendicularly to the current (ICI case). This case, where the dominant component is an equatorial dipole, strongly reminds  the geometry of the magnetic field that was numerically generated by using the mean flow measured in a VK geometry with counter-rotating propellers \cite{marie2003,bourgoin2004}. This emphasizes the similarity between VK and TG flows. 
Note however that the mean flow component in the VK geometry is axisymmetric and cannot drive an axisymmetric magnetic field because of Cowling's theorem. A magnetic field with an equatorial dipolar component provides a simple way to break axisymmetry. In contrast, the TG forcing is not axisymmetric and thus does not enforce such a strong constraint on the generated magnetic field. Indeed, when symmetry constraints related to infinite magnetic permeability are implemented for all boundaries (III case), the generated magnetic field involves a dominant component which is an axial dipole.  This corresponds to the geometry of the magnetic field generated in the VKS experiment, where it was ascribed to the presence of non-axisymmetric velocity fluctuations \cite{petrelis2007}. 

Direct numerical simulations of dynamos generated by an axisymmetric $s2-t2$ forcing in a sphere, a configuration similar to the VK forcing, have shown that an equatorial dipole is observed at low ${\rm Re}$, for which the flow is axisymmetric, whereas an axial dipole is obtained at higher ${\rm Re}$, i.e. in the presence of non-axisymmetric velocity fluctuations \cite{gissinger2008}. We observe that the TG forcing provides a different scenario: the axial dipole does not seem to be favored by turbulent fluctuations (compare Fig. \ref{Fig:sigmaBsym30} a and b). Thus, slight deviations from axisymmetry, present in the TG forcing, are enough to generate a magnetic field with a dominant axial dipolar component, even at small ${\rm Re}$, provided appropriate boundary conditions are simulated. 

Finally, we have studied the nonlinear saturation of the axial dipolar magnetic field (III case). We have observed that the dynamo bifurcation can be subcritical in some parameter range. We have shown that this phenomenon can be explained by the presence of an hydrodynamic instability that competes with the growing magnetic mode. This is not unlikely for ${\rm P_m}$ of order $1$ which is the case of most numerical simulations and provides a general mechanism for subcritical dynamos. It would be of interest to check whether the subcritical dynamo bifurcations reported previously in non confined TG flows \cite{pontylaval2007}, 
flows forced in a sphere \cite{fuchs1999} or simulations of the geodynamo  \cite{Christensen}, are also related to a similar mechanism.
Other open questions concern the dynamics of the magnetic field above the dynamo threshold. Breaking the symmetry of the TG flow in order to generate reversals of the axial dipole or other dynamics resulting from nonlinear coupling between different magnetic modes deserve to be studied.

\textbf{Acknowledgments:} 
The computations were carried out at CEMAG  and IDRIS (CNRS).

\bibliographystyle{apsrev}

\end{document}